\documentclass[natbib,twocolumn]{svjour3}          

\usepackage{mathptmx}

\usepackage[utf8]{inputenc}

\usepackage{amsmath}
\usepackage{amsfonts}
\usepackage{amssymb}

\usepackage{graphicx}
\graphicspath{ {./graphics/} }
\usepackage{url}

\usepackage{xcolor}
\usepackage{booktabs}
\usepackage{arydshln}
\usepackage{balance}

\usepackage[hidelinks]{hyperref}     

\usepackage[ruled,lined,linesnumbered]{algorithm2e}

\def\makeheadbox{\relax}

\def \calG {\mathcal{G}}
\def \calN {\mathcal{N}}
\def \calE {\mathcal{E}}
\def \calL {\mathcal{L}}

\newcommand{\mtrx}[1]{{#1}}
\newcommand{\vect}[1]{{#1}}

\begin{document}

\title{Inference of Demographic Attributes based on
  Mobile Phone Usage Patterns and Social Network Topology}

\author{Carlos Sarraute
 \and Jorge Brea
 \and Javier Burroni
 \and Pablo Blanc
 }

\institute{
 C. Sarraute \and J. Brea \and J. Burroni
 \at
 Grandata Labs \\
 Bartolomé Cruz 1818, Vicente Lopez \\
 Buenos Aires, Argentina \\
 {e-mail: charles@grandata.com}
\and
 J. Brea \at
 {e-mail: jorge@grandata.com}
\and
 J. Burroni \at
 {e-mail: javier.burroni@grandata.com} 
\and
 P. Blanc 
 \at
 IMAS, UBA-CONICET.\\
 FCEN, Ciudad Universitaria,\\
 Int Guiraldes 2160, CABA, Argentina.\\
 {e-mail: pblanc@dm.uba.ar}
 }

\date{First Online: 12 July 2015}

\maketitle

\begin{abstract}

Mobile phone usage provides a wealth of information,
which can be used to better understand the demographic
structure of a population. In this paper we focus on the
population of Mexican mobile phone users.
We first present an observational study of mobile phone
usage according to gender and age groups.
We are able to detect significant differences in phone usage among 
different subgroups of the population.
We then study the performance of different machine learning methods to
predict demographic features (namely age and gender) of unlabelled
users by leveraging individual calling patterns, as well as the
structure of the communication graph. We show how a specific implementation 
of a diffusion model, harnessing the graph structure, has significantly 
better performance over other node based standard machine learning methods. 
We provide details of the methodology together with an analysis of the robustness of our
results to changes in the model parameters. Furthermore, 
by carefully examining the topological
relations of the training nodes (seed nodes) to the rest of the nodes in the network, 
we find topological metrics which have a direct influence on the performance
of the algorithm. 

\keywords{
  Social network analysis 
  \and Mobile phone social network
  \and Call Detail Records  
  \and Big Data 
  \and Graph mining  
  \and Demographics 
  \and Homophily
  \and Semi-supervised learning 
}
\end{abstract}

\section{Introduction}

Mobile phones have become prevalent in all parts of the world,
in developed as well as developing countries, and
provide an unprecedented source of information on the 
dynamics of the population on a national scale.
This opened the way to a very active field of research.
For instance, \cite{gonzalez2008understanding} is one of the first works which analysed trajectories 
of mobile phone users in order to model human mobility, uncovering spatial and temporal regularities,
such as the tendency of users to return to a few highly frequented locations.
Such mobility model also allowed the study of viral dynamics in mobile phone networks
in \cite{wang2009understanding},
characterizing the spreading patterns of a mobile virus.
More recently, \cite{dyagilev2013information} have analysed information propagation  
on mobile phone networks,
and \cite{naboulsi2015mobile} provided a survey of
the multidisciplinary activities that rely on mobile traffic datasets.
In particular, mobile phone usage is being used to
perform quantitative analysis of the demographics of a population,
respect to key variables such as gender, age, level of education
and socioeconomic status -- for example see the works of \cite{blumenstock2010mobile}
and \cite{blumenstock2010s}.
Other key aspects of mobile phone networks is their
growth and structure \citep{perc2014matthew},
and the fact that their complexity is better modeled by 
multilayer networks \citep{kivela2014multilayer} rather than ordinary graphs.

In this work we combine two sources of information: 
transaction logs from a major mobile phone operator,
and information on the age and gender of a subset of
the mobile phone user population.
This allows us to perform an observational study of 
mobile phone usage, differentiated by gender and age groups.
This study is interesting in its own right, 
since it provides knowledge on the structure and
demographics of the mobile phone market in the studied country.
We can start to fill gaps in our understanding of basic demographic questions:
Are differences between men and women, as reported by \citet{katz_economics_2001},
reflected in mobile phone usage (in calling and texting patterns)?
What are the differences in mobile phone usage between distinct age ranges?
 
The second contribution of this work is to apply 
the knowledge on calling patterns to predict demographic features, 
namely to predict the age and gender of unlabeled users.
This is, we want to infer demographics from social behavior -- for example see \citep{adali2014predicting} where personality traits are inferred. 
We present methods that rely on individual calling patterns,
and introduce a diffusion based algorithm that exploits the structure of the social graph (induced by communications),
in order to improve the accuracy of our predictions.

Being able to understand and predict demographic features such as age and
gender has numerous applications, 
from market research and segmentation to the 
possibility of targeted campaigns, such as health campaigns for women \citep{frias2010gender}.

The remainder of the paper is organized as follows:
in Sect.~\ref{sec:exploratory} we provide an overview of the datasets used in 
this study. 
Section~\ref{sec:observations} describes the observations that we gathered,
the insights gained from data analysis, and
the differences that could be seen in CDR features between genders and age groups.
In particular, very clear correlations have been observed in the links between users
according to their age.
In Sect.~\ref{sec:identification} we present the models that we used to identify
the age and gender of unlabeled users.
We show the experimental results obtained using classical Machine Learning techniques
based on individual attributes, both for gender and age.
We then introduce in Sect.~\ref{sec:age-identification-links} an algorithm that 
leverages the links between users using only
the graph topology, and then combine it with standard machine learning techniques.
In our experiments the pure graph based algorithm has the best predictive power.
In Sect.~\ref{relatedwork} we review related work and state how our results augment this body 
of work. 
Section~\ref{sec:conclusion} concludes the paper with a summary of our results and ideas for future work.

\section{Dataset Exploratory Analysis} \label{sec:exploratory}

In this section we present our dataset, and 
report on the exploratory quantitative analyses that we performed.
Our raw data input are the transaction logs (that contain billions of records at the 
scale of a country). A first step in the study was 
to generate characterization variables for each user, which
summarize their individual and social behavior.
We also describe the preprocessing performed on the data,
and the key features identified with 
PCA (Principal Component Analysis).

\subsection{Dataset Description} \label{sec:dataset}

The dataset used for this study consists of cell phone call 
and SMS (Short Message Service) records
collected for a period of $M$ months ($M = 3$) by a large mobile phone operator.
The dataset is anonymized.
For our purposes, each CDR (Call Detail Record) is represented as a tuple 
$\left < x, y, t, \Delta, d, l \right >$,
where $x$ and $y$ are the encrypted phone numbers of the caller and the callee,
$t$ is the date and time of the call,
$\Delta$ is the duration of the call,
$d$ is the direction of the call (incoming or outgoing, with respect to the mobile operator client),
and $l$ is the location of the tower that routed the communication.
Similarly, each SMS record is represented as a tuple $\left < x, y, t, d \right >$.

We construct a social graph $\calG = \left < \calN, \calE \right > $,
based on the aggregated traffic of $M$ months.
We use $\calN$ to denote the set of mobile phone users that appear in the dataset.
$\calN$ contains about 90 million unique cell phone numbers.
Among the numbers that appear in $\calN$, only some of them are clients of the 
mobile phone operator: we denote that set $\calN_O$.

For this study, we had access to basic demographic information
for a subset of the nodes (clients and non clients of the mobile operator), that we denote $\calN_{GT}$ 
(where \emph{GT} stands for \emph{ground truth}). This data was provided by the mobile phone operator but we have no information or control on how this subset of users was selected.
The size of this labeled set $| \calN_{GT} | $ is over 500,000 users.
The following relations hold between the three sets: $\calN_{GT} \subset \calN$
and $\calN_O \subset \calN$.

\begin{figure}[t]
	\centering
	{\includegraphics[trim=1.2cm 0.3cm 1.2cm 1.2cm, clip=true, width=0.95\linewidth]{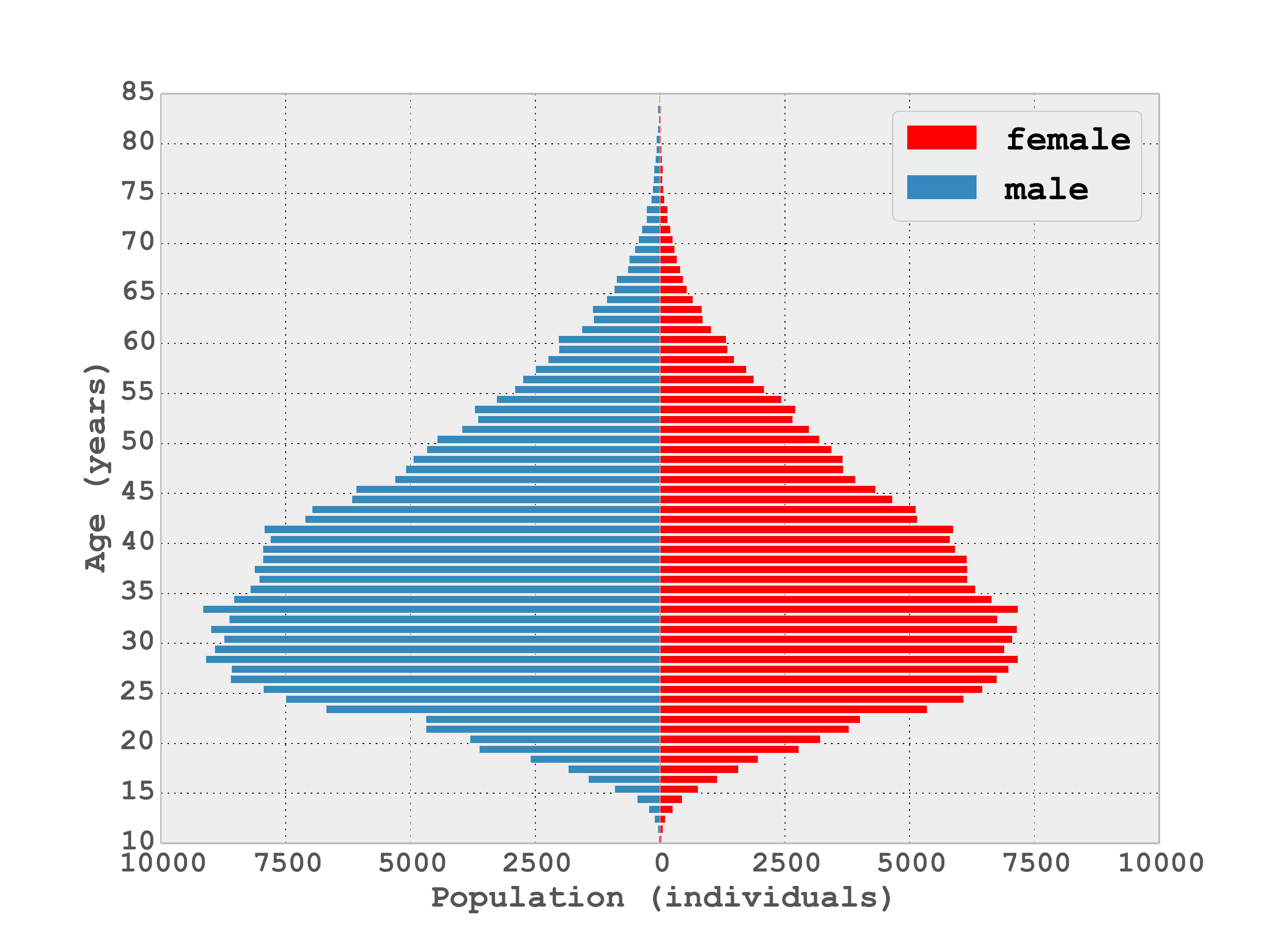}}
	{\includegraphics[trim=1.2cm 0.3cm 1.2cm 1.2cm, clip=true, width=0.95\linewidth]{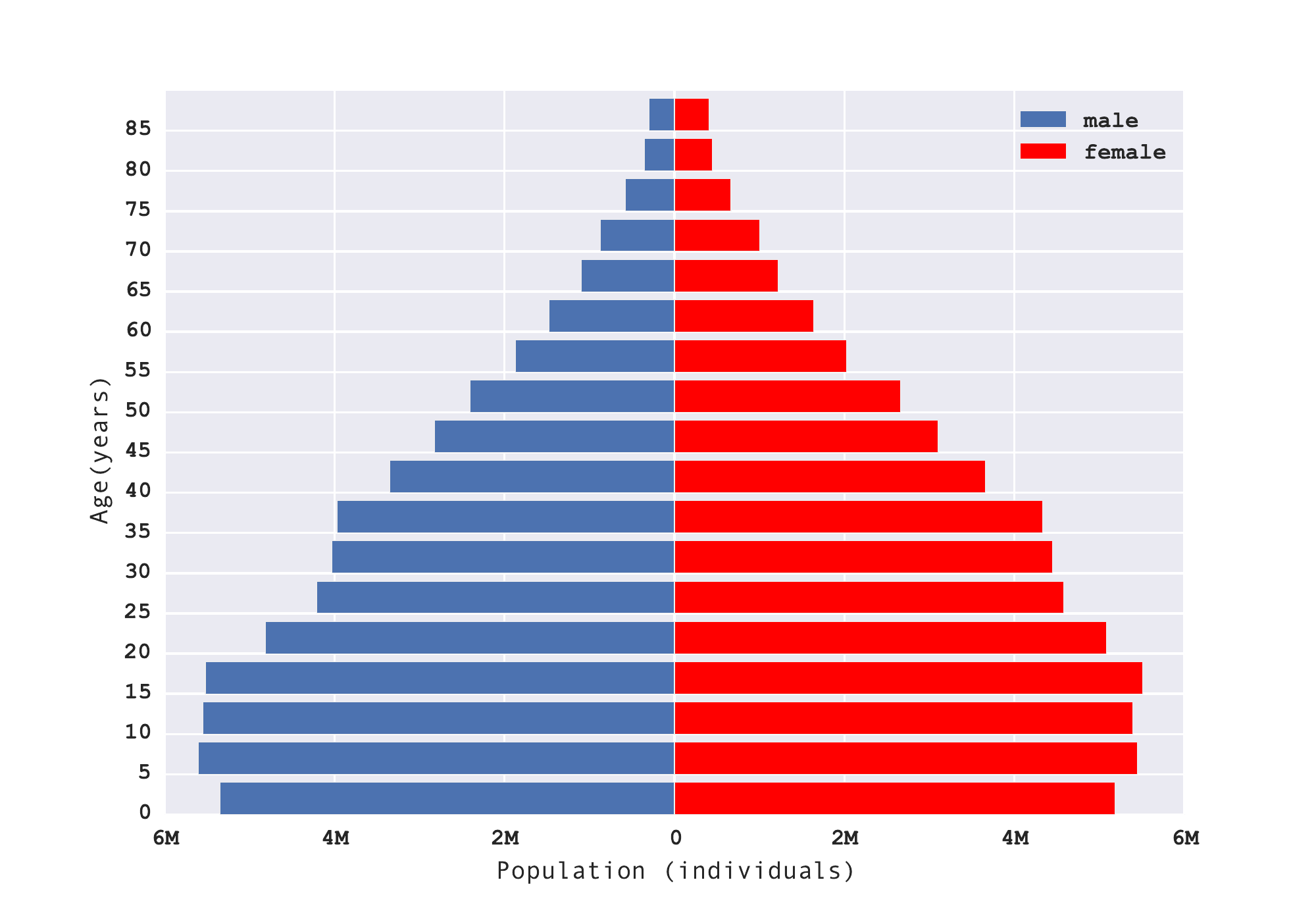}}
	\caption{Pyramid of ages, discriminated by gender, of the labeled set $\calN_{GT}$, showing the number of
	individuals for each age value (top). Pyramid of ages, discriminated by gender, of the whole population taken from census data (bottom). }
	\label{fig:pyramid}
\end{figure}

Fig.~\ref{fig:pyramid}(top) shows the pyramid of ages of the labeled set.
This pyramid is different from the age pyramid of the entire population (shown in Fig.~\ref{fig:pyramid}(bottom)),
since it only contains mobile phone users (that pertain to $\calN_{GT}$). 
The data of Fig.~\ref{fig:pyramid}(bottom) comes from the National Institute of Statistics and Geography.
The difference between both pyramids is most noticeable in the age range below 35 years old.
Since all the nodes in our graph are mobile phone users (belonging to $\calN$),
we expect their demographics to be better described by Fig.~\ref{fig:pyramid}(top). 

Some basic observations on $\calN_{GT}$: there are more men ($56.83\%$)
than women ($43.17\%$) in the labeled set.
The mean age is $37.23$ years for men and $36.47$ years for women.
On the other hand, in the whole population, there 
are slightly more women ($51.2\%$) than men ($48.8\%$);
and the mean age is $28.9$ years for men and $30.0$ years for women.

\subsection{Characterization Variables} \label{sec:variables}

For users in $\calN$, we computed the following variables which characterize their
calling consumption behavior (also called ``behavioral variables" by \citealt{frias2010gender}).

\begin{itemize}
\item \textit{Number of Calls}. We consider incoming calls i.e.,
the total number of calls received by user $u$ during a period
of three months, as well as outgoing calls i.e., total number of
calls made by user $u$.
Additionally, we distinguish whether those calls happened
during the weekdays (Monday to Friday) or during the weekend; 
and we further split the weekdays in two parts: the ``daylight'' (from 7 a.m. to 7 p.m.)
and the ``night'' (before 7 a.m. and after 7 p.m.).

We thus have $3 \times 4$ variables for the number of calls, given by the
Cartesian product: [ in, out, all ] $\times$ [weekdaylight, weeknight, weekend, total].

\item \textit{Duration of Calls}. We calculate the total duration 
of incoming calls and outgoing calls of user $u$ during the period of three months. 
As before, we distinguish between weekdays (by daylight and by night) and weekends,
to get a total of 12 variables for the duration of calls.

\item \textit{Number of SMS}. We consider incoming messages 
(received by user $u$)
and outgoing messages (sent by user $u$).
Similarly we distinguish between weekdays (by daylight and by night) and weekends,
to get a total of 12 variables for the number of SMS.

\item \textit{Number of Contact Days}. We consider the number of days where the user has activity. We distinguish between calls and SMS, and between incoming, outgoing or any activity. This way we get 6 variables related to the number of activity days. 

\end{itemize}

We also computed variables which characterize the social network of users
based on their use of the cell phone (also called ``social variables'' by \citealt{frias2010gender}).

\begin{itemize}
\item \textit{In/Out-degree of the Social Network}. The in-degree for
user $u$ is the number of different phone numbers
that called or sent an SMS to that user. 
The out-degree is the number of distinct phone numbers
contacted by user $u$.

\item \textit{Degree of the Social Network}. 
The degree is the number of unique phone numbers
that have either contacted or been contacted by user
$u$ (via voice or SMS).

\end{itemize}

\subsection{Data Preprocessing} \label{sec:preprocessing}

Many of the variables that we generated have a right skewed or heavy tailed distribution.
Our experiments showed that this skewness affects the results
given by the Machine Learning algorithms we describe in Sect.~\ref{sec:identification}.
Therefore as part of the data preprocessing we also considered the logarithmic version of each variable.
We discuss this preprocessing in more detail for one variable, that we 
use as running example: \emph{in-time-total}, i.e. the total duration of incoming calls
for a given user.

\begin{table}[t]
\caption{Statistic summary for \textit{in-time-total} and its logarithmic transformation.}
\label{tab:prelog}
\setlength{\tabcolsep}{11pt}

\begin{tabular}{ l r r }
\toprule
{}    &  \textit{in-time-total} (seconds) & $\log(\textit{in-time-total} + 1)$ \\
\midrule
count &   131770.00 &  131770.00 \\
mean  &    16239.28 &       3.31 \\
std   &    50023.16 &       1.23 \\
min   &        0.00 &       0.00 \\
25\%  &      662.00 &       2.82 \\
50\%  &     3838.00 &       3.58 \\
75\%  &    14108.00 &       4.14 \\
max   &  4045686.00 &       6.60 \\
\bottomrule
\end{tabular}

\end{table}

\begin{figure}[t]
	\centering
	{\includegraphics[clip, trim=0cm 0cm 0.9cm 0.4cm, width=0.9\linewidth]{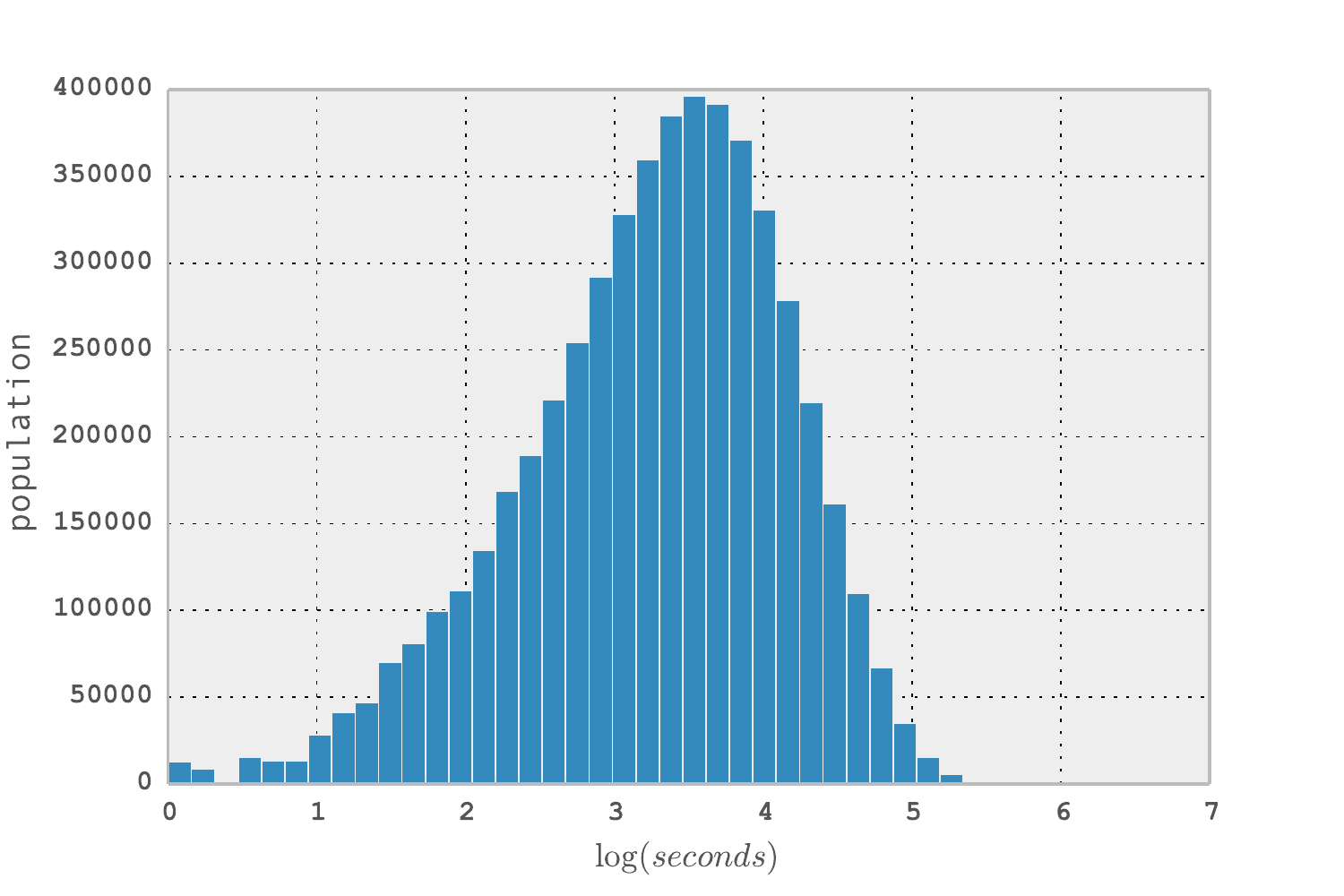}}
	\caption{Histogram of $\log(\textit{in-time-total+1})$. 
The distribution has roughly a bell shape for $\log(\textit{in-time-total} + 1) > 0$.
}
	\label{fig:logintimetotal}
\end{figure}

As can be seen in Table~\ref{tab:prelog} (left), the quartiles for the variable \emph{in-time-total} 
lie in different orders of magnitude, 
in particular the ratio ${IQR}/{Q_2} = {(Q_3-Q_1)}/{Q_2}$ is well above $1$,
which is a characteristic of right skewed distributions.\footnote{The interquartile range ($IQR$) is a measure of statistical dispersion. The ratio ${IQR}/{Q_2}$ is the non-parametric version of the Coefficient of Variation.}
To improve the results given by the Machine Learning methods,
we transform the data using the function
$ T(x) = \log_{10}(x + 1) $.
After the transformation, we found the statistics in Table~\ref{tab:prelog} (right). 
As we can see, the quartiles are in the same order of magnitude, 
and the ratio ${IQR}/{Q_2} $ is below $1$. 
The resulting distribution is shown in Fig.~\ref{fig:logintimetotal}.

In conclusion, we decided to include both plain variables as well as their logarithmic values,
and let our Machine Learning algorithms select which variables are most relevant
for modeling a given target variable (e.g. gender and age).
We also rescaled all variables to values between 0 and 1.

\subsection{Insights on Key Features from PCA} \label{sec:pca}

We performed PCA (Principal Component Analysis) on the behavioral variables,
in order to gain information on which are the most important variables.
This gave us interesting insights on key features of the data.
We describe the first 4 eigenvectors (which account for $89.6\%$ of the variance).

The first eigenvector retains $76.0\%$ of the total variance.
This eigenvector is dominated by the logarithmic 
version\footnote{We note that in all eigenvectors, the logarithmic version of the variables got systematically higher coefficients than the plain variables;
which is expectable since they have higher variance.}
of the total number of calls, total duration of calls and total number of SMS.
This result shows that the level of activity of users exhibits the highest variability,
and therefore is a good candidate to characterize users' social behavior.

The second eigenvector, which retains $6.7\%$ of the variance,
gives high positive coefficients to ``outgoing'' variables
(number of outgoing calls, duration of outgoing calls, number of SMS sent) and negative coefficients to  
``incoming'' variables (number of incoming calls, duration of incoming calls, number of SMS received). 
This suggests that the difference of outgoing minus incoming 
communications is also a good variable to describe users' social behavior.

The third eigenvector (with $4.4\%$ of the variance) gives positive coefficients
to the ``voice call'' variables, and negative coefficients to the ``SMS'' variables
(intuitively the difference between voice and SMS usage is relevant).

The fourth eigenvector (with $2.5\%$ of the variance) gives positive coefficients
to the ``weeknight'' and ``weekend'' variables (communications made non-working hours, i.e. during the night or during the weekend), and negative coefficients to ``weeklight'' variables (communications made during the day, from Monday to Friday, which correspond roughly to working hours).

It is interesting to note that our PCA analysis selected as eigenvectors with strongest variance
a set of variables which also have a clear semantic interpretation: 
they define behavioral characteristics of the individuals.

\section{Observational Study} \label{sec:observations}

In this section we present the insights that we gained
from the exploratory analysis.
We found statistically significant differences with respect to gender and age,
which motivate our attempt to identify these attributes based on the communication patterns
(we describe the inference algorithms in Sect.~\ref{sec:identification}
and \ref{sec:age-identification-links}).
We will focus on some examples to illustrate the kind of observations we obtained for many variables.

\subsection{Observed Gender Differences} \label{sec:observations-gender}

We report in Table~\ref{tab:mean-male-female} the mean value for 6 key variables, 
for both genders. 
We know from PCA that the number of calls,
and the total duration of calls made by a user, characterize their level of activity.
A natural question is whether differences can be observed between genders.
We found that for all the variables, there is a significant difference between genders
($p$-value $< 10^{-10}$).
Recall that these values are computed as the aggregation of calls during a period
of $M = 3$ months for users in $\calN_{GT}$ (about 500,000 users).

\begin{table}[t]
\caption{Sample mean of key variables for female and male users. 
The durations are expressed in seconds.
}
\label{tab:mean-male-female}
\setlength{\tabcolsep}{10pt}

\begin{tabular} { p{0.45 \linewidth} r r }
\toprule
Variable & Female & Male \\
\midrule
$\hat{\mu}$(mean duration)          & 103.64 & 104.95 \\
$\hat{\mu}$(mean duration outgoing) &  100.52 &  107.57 \\
$\hat{\mu}$(mean duration incoming) &  103.59 &  99.52 \\
\midrule
$\hat{\mu}$(number of calls total)    & 72.847  & 81.348 \\
$\hat{\mu}$(number of calls outgoing) & 44.136  & 50.047 \\
$\hat{\mu}$(number of calls incoming) & 28.710  & 31.301 \\
\bottomrule
\end{tabular}

\end{table}

Table~\ref{tab:mean-male-female} shows that men have on average
higher total number of calls, and mean duration of calls (measured in seconds per call).
However an interesting pattern can be seen when we distinguish incoming
and outgoing calls: the mean duration of \emph{outgoing} calls is higher for men,
but the mean duration of \emph{incoming} calls is higher for women.
It follows that the net mean duration of calls (the difference between mean outgoing and incoming calls)
has a marked gender difference: 
the sample mean $\hat{\mu}$(net total duration) = 8.05 seconds per call for men,
and $\hat{\mu}$(net total duration) = -3.07 seconds per call for women.\footnote{We note that the number of outgoing calls is higher than the number of incoming calls
for both men and women, due to a particularity of our dataset
(for all the users in $\calN$ the total number of incoming and outgoing calls is
the same, but for users in $\calN_{GT}$ there is a higher proportion of outgoing calls).}

\begin{figure}[t]
\centering
\begin{footnotesize}
	\begin{minipage}{.48\linewidth}
		\centering
		\includegraphics[width=\textwidth]{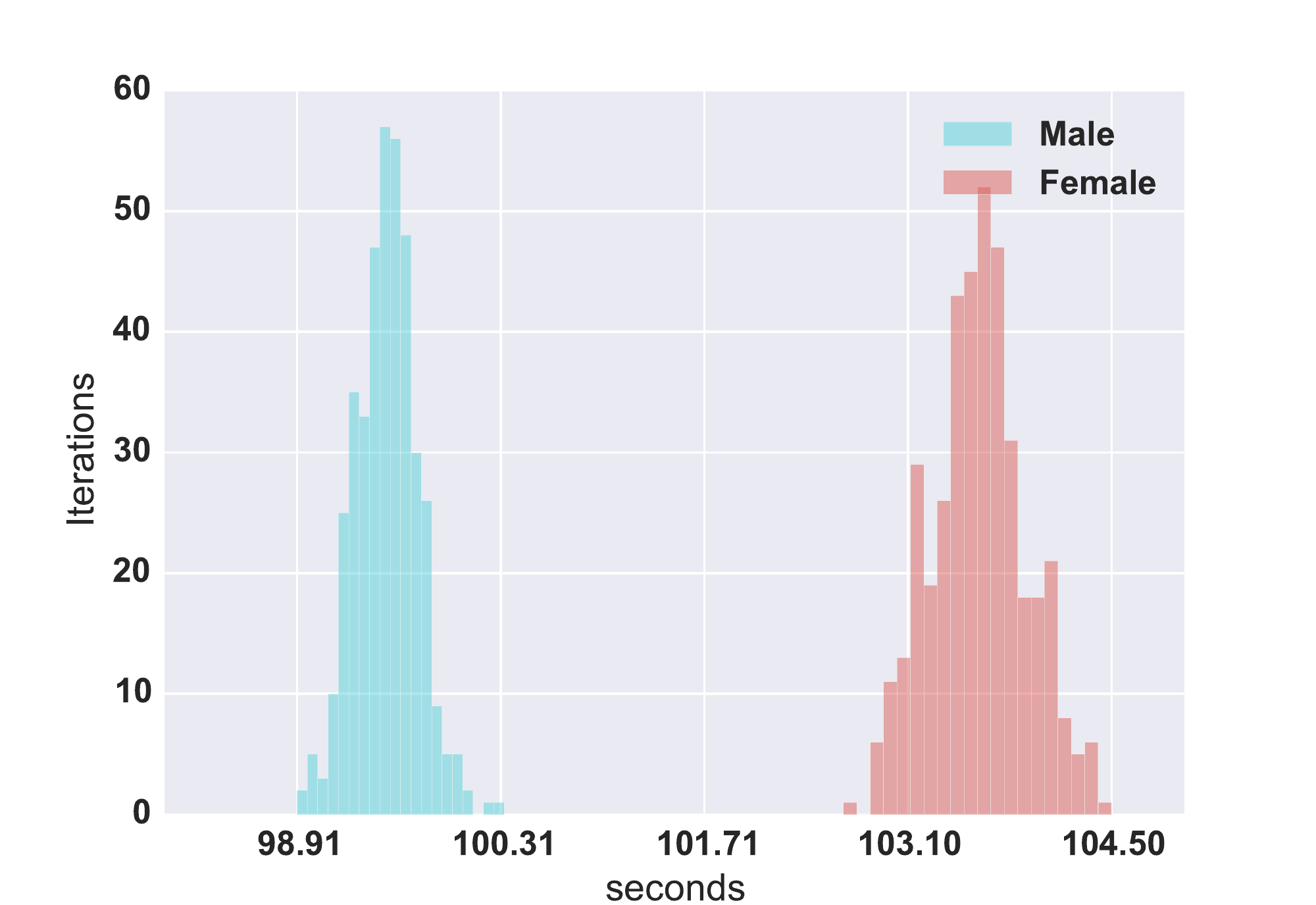}
		(a) 
	\end{minipage}
	\begin{minipage}{.48\linewidth}
		\centering
		\includegraphics[width=\textwidth]{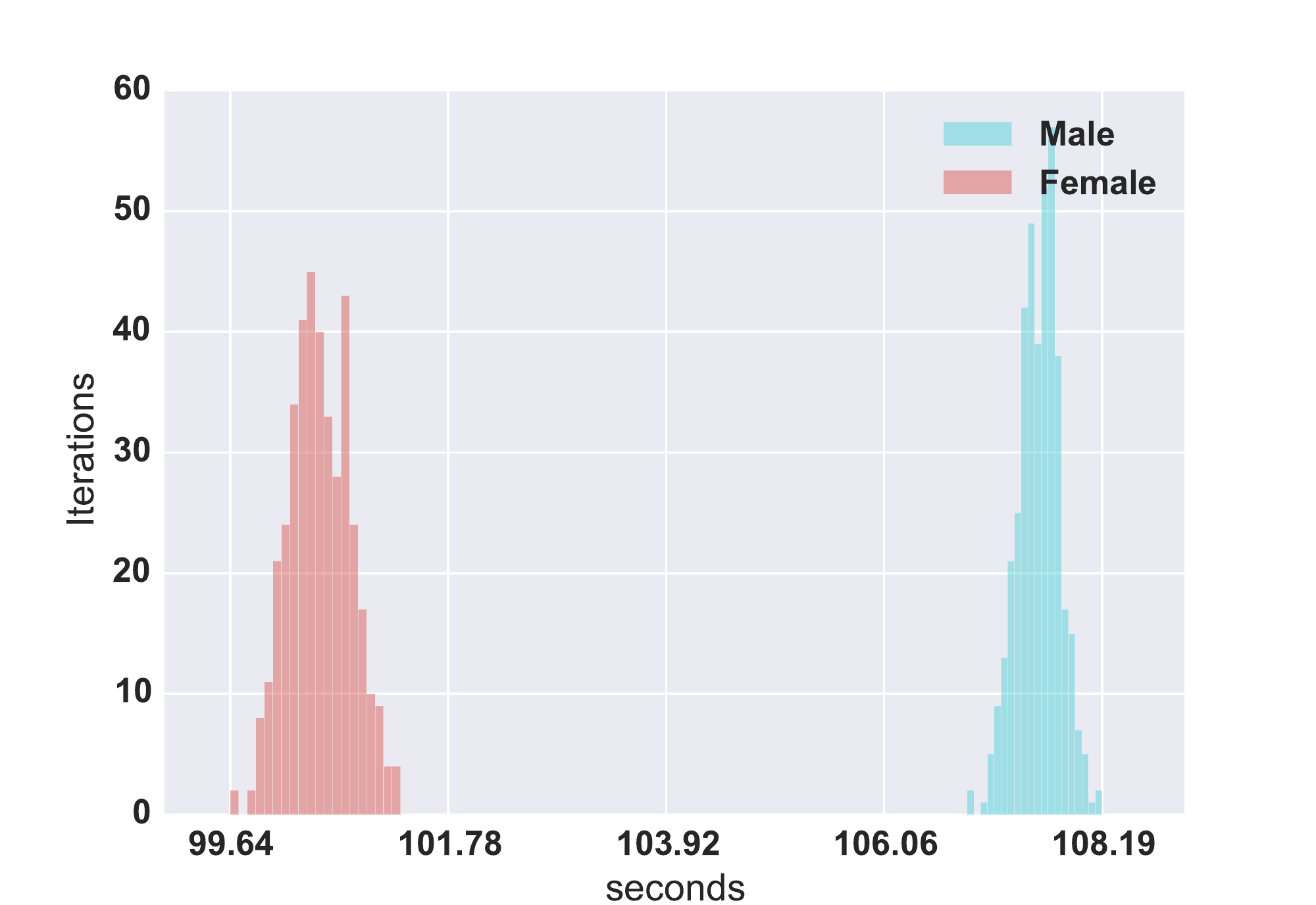}
		(b) 
	\end{minipage}
	\begin{minipage}{.48\linewidth}
		\centering
		\includegraphics[width=\textwidth]{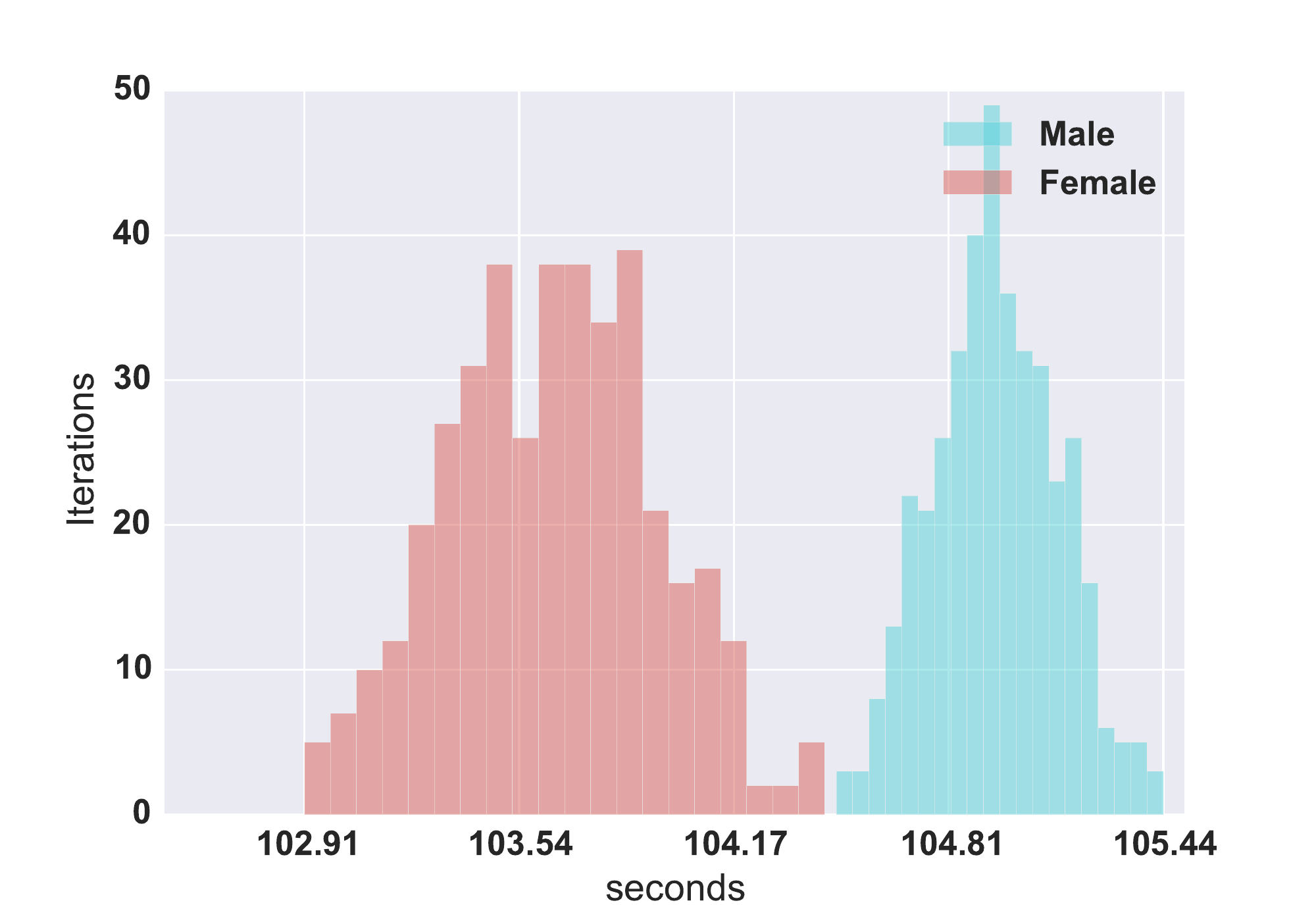}
		(c) 
	\end{minipage}
	\begin{minipage}{.48\linewidth}
		\centering
		\includegraphics[width=\textwidth]{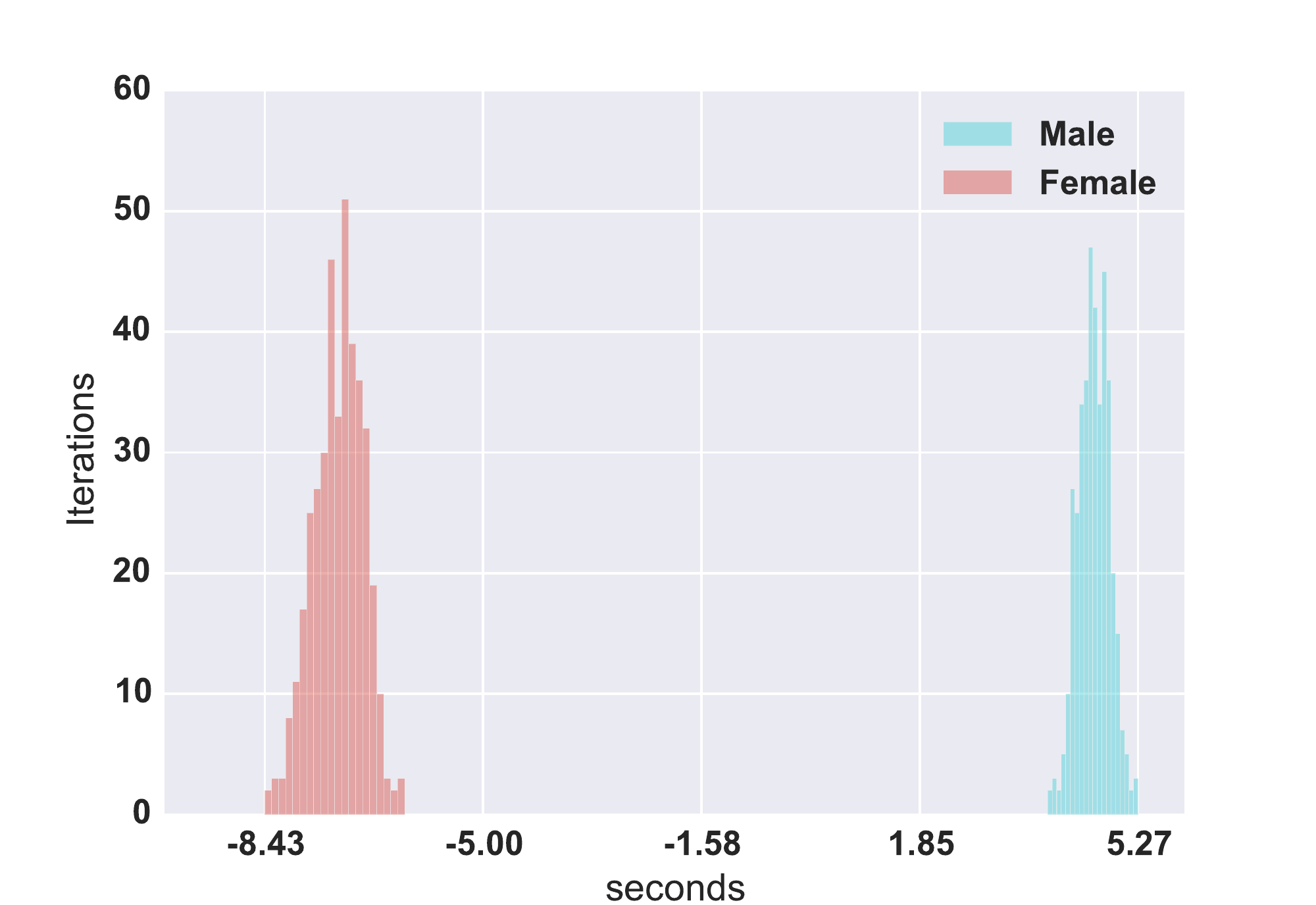}
		(d) 
	\end{minipage}
\end{footnotesize}

\caption{Bootstrapping results: (a) incoming calls, (b) outgoing calls, (c) all calls, (d) net duration. We can see significant differences for both incoming and outgoing call duration between genders. Men make longer outgoing calls, while woman make longer incoming calls}
\label{fig:bootstrapping}
\end{figure}

To have a better test of the statistical significance of these results, we generated a set of 400 new population samples for both incoming and outgoing calls using the  bootstrapping method where each new population sample is generated by randomly selecting $m$ values with replacement from the original population sample, where $m$ is equal to the size of the original population sample. This allows us to construct distribution of mean values for each of our variables of interest, and thus compare distributions instead of a single point value. We note that we are building the distribution for the variable's  \textbf{mean} value, and not the value of the variable itself. 
In Fig.~\ref{fig:bootstrapping}(a) and \ref{fig:bootstrapping}(b) we show the distributions of the \textbf{mean} incoming and outgoing calls for both genders, where we can clearly see that these two distributions are not even overlapping. Furthermore, we plot the distributions for the total number of calls for each gender in Fig.~\ref{fig:bootstrapping}(c) and the net duration of calls\footnote{For the net duration of calls we considered only users which had both incoming and outgoing calls.} in Fig.~\ref{fig:bootstrapping}(d), again the distributions of the mean values for each gender are non overlapping. Notice these results do not imply that gender are easily identifiable, instead, they show that the difference between their means are significant, without  making any assumptions on their underlying distribution.

We also computed the conditional probability $p(g'|g)$ that a random call made by an 
individual with gender $g$ has a recipient with gender $g'$, 
where we denote male by $M$ and female by $F$. 
For the calls originated by male users, we found that 
$p(F|M) = 0.3735$ and $p(M|M) = 0.6265$. 
For the calls originated by female users, we found that
$p(F|F) = 0.4732$ and $p(M|F) = 0.5268$.
We can see a difference between genders, 
in particular men tend to talk more with men, and women tend to talk more with women.
More precisely:
\begin{equation}
\label{eq:gender-homophily}
\begin{split}
p(M|F) < p(M) &= 0.5683 < p(M|M) \\
p(F|M) < p(F) &= 0.4317 < p(F|F)
\end{split}
\end{equation}
Similar observations have been made in the case of the Facebook social graph~\citep{ugander2011anatomy}.

\subsection{Observed Age Differences} \label{sec:observations-age}

We approached the study of mobile phone usage patterns according to age
by dividing the population in $C=4$ categories: 
below 25, from 25 to 34, from 35 to 49, and above 50 years old. 

Since we are dealing with more than $2$ groups, comparing differences between groups 
requires using the correct tool, as the probability of making a type I error (null hypothesis incorrectly rejected) increases. 
In order to compare the means (of the $\log$) of the variables for each age group, 
we conduct a Tukey's Honest Significant Difference (HSD) test \citep{tukey1949comparing}.
This method tests all groups, pairwise, simultaneously.  
We found a list of 20 variables for which the null hypothesis of same mean ($H_0$) is rejected for all pair of groups, i.e. $\mu_i \neq \mu_j$ for every $i \neq j$.

\begin{table}[t]
\caption{Tukey HSD for the variable $\log_{10}$(\textit{in-time-total} + 1)
The 4 age groups are labeled 0, 1, 2, 3. Pairwise comparisons are done for all combinations (of group1 and group2). The null hypothesis is rejected for all pairs;
in other words, all the groups are found statistically different respect to this variable.
}
\label{tab:tukeyHSD}
\setlength{\tabcolsep}{7pt}

\begin{tabular}{ c c c c c c }
\toprule
group1 & group2 & meandiff & lower & upper & reject  \\
\midrule
       0        &        1        &       0.1567      &     0.1328     &     0.1807     &       True       \\
       0        &        2        &       0.1326      &     0.1088     &     0.1564     &       True       \\
       0        &        3        &       0.2367      &     0.2122     &     0.2612     &       True       \\
       1        &        2        &      -0.0242      &    -0.0407     &    -0.0076     &       True       \\
       1        &        3        &       0.0800       &     0.0625     &     0.0975     &       True       \\
       2        &        3        &       0.1041      &     0.0868     &     0.1214     &       True       \\
\bottomrule
\end{tabular}

\end{table}

\begin{figure}[t]
\centering
\begin{footnotesize}

	\begin{minipage}{.48\linewidth}
		\centering
		\includegraphics[width=\textwidth]{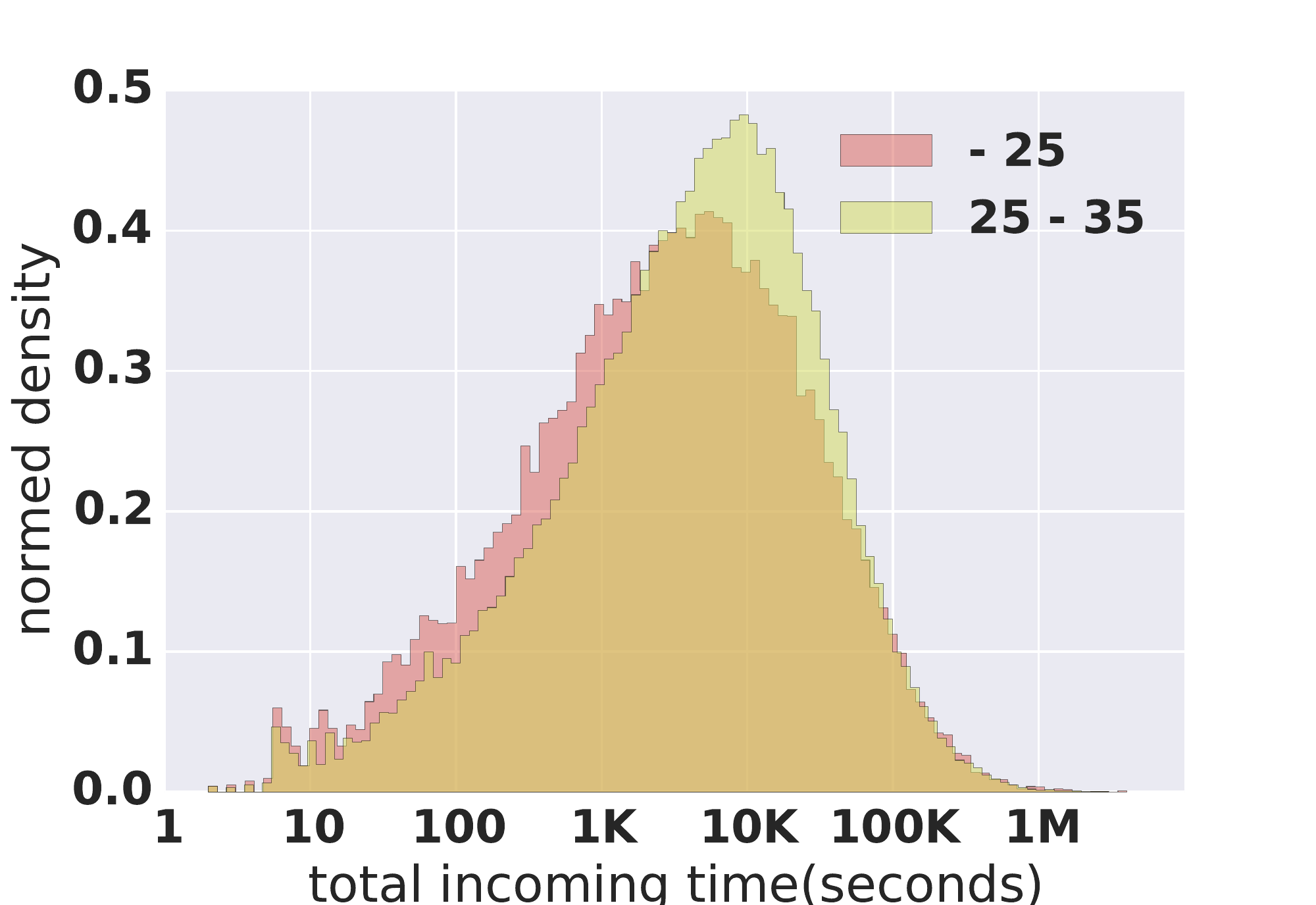}
		(a) 
	\end{minipage}
	\begin{minipage}{.48\linewidth}
		\centering
		\includegraphics[width=\textwidth]{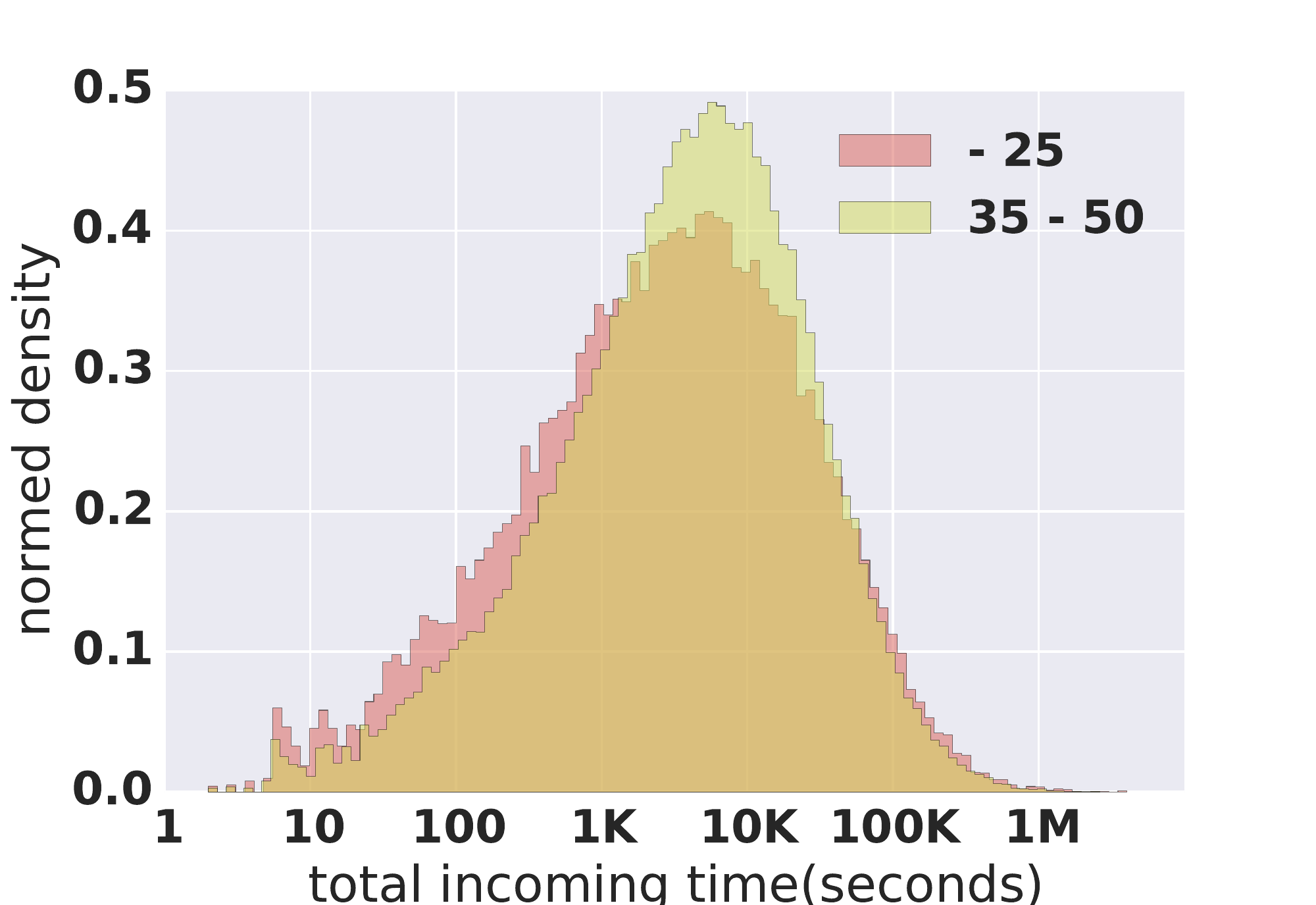}
		(b) 
	\end{minipage}
	\begin{minipage}{.48\linewidth}
		\centering
		\includegraphics[width=\textwidth]{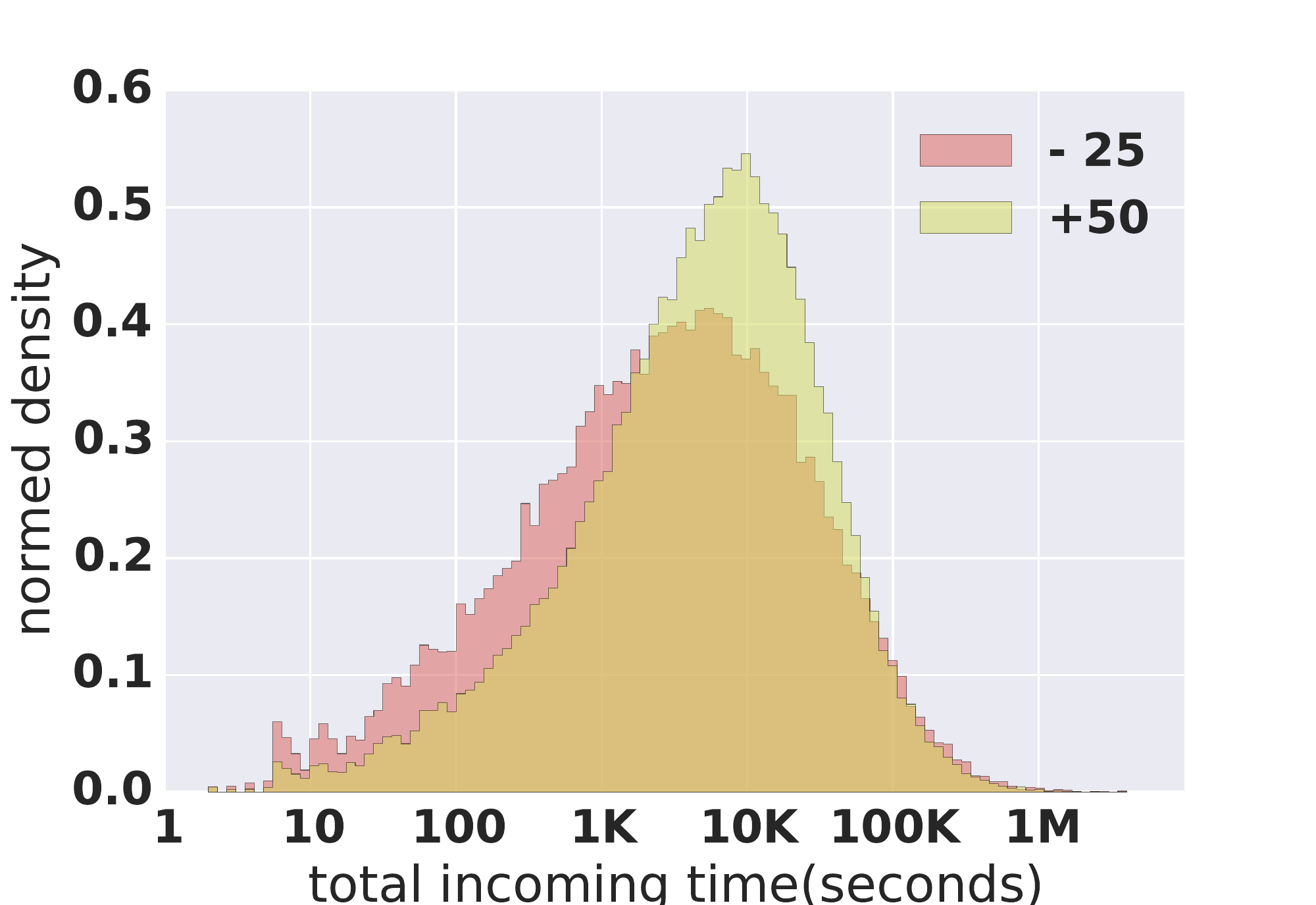}
		(c) 
	\end{minipage}
	\begin{minipage}{.48\linewidth}
		\centering
		\includegraphics[width=\textwidth]{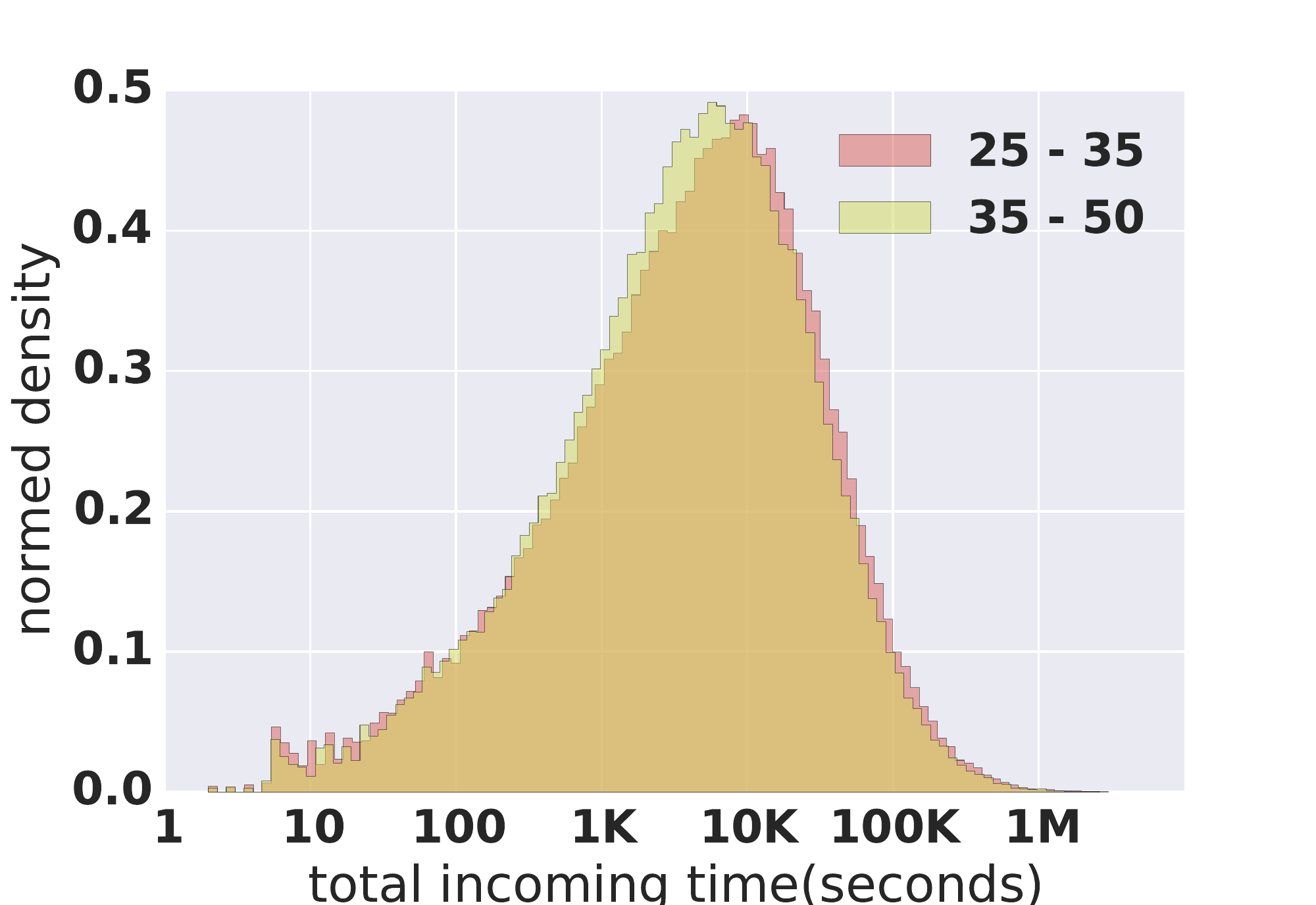}
		(d) 
	\end{minipage}
	\begin{minipage}{.48\linewidth}
		\centering
		\includegraphics[width=\textwidth]{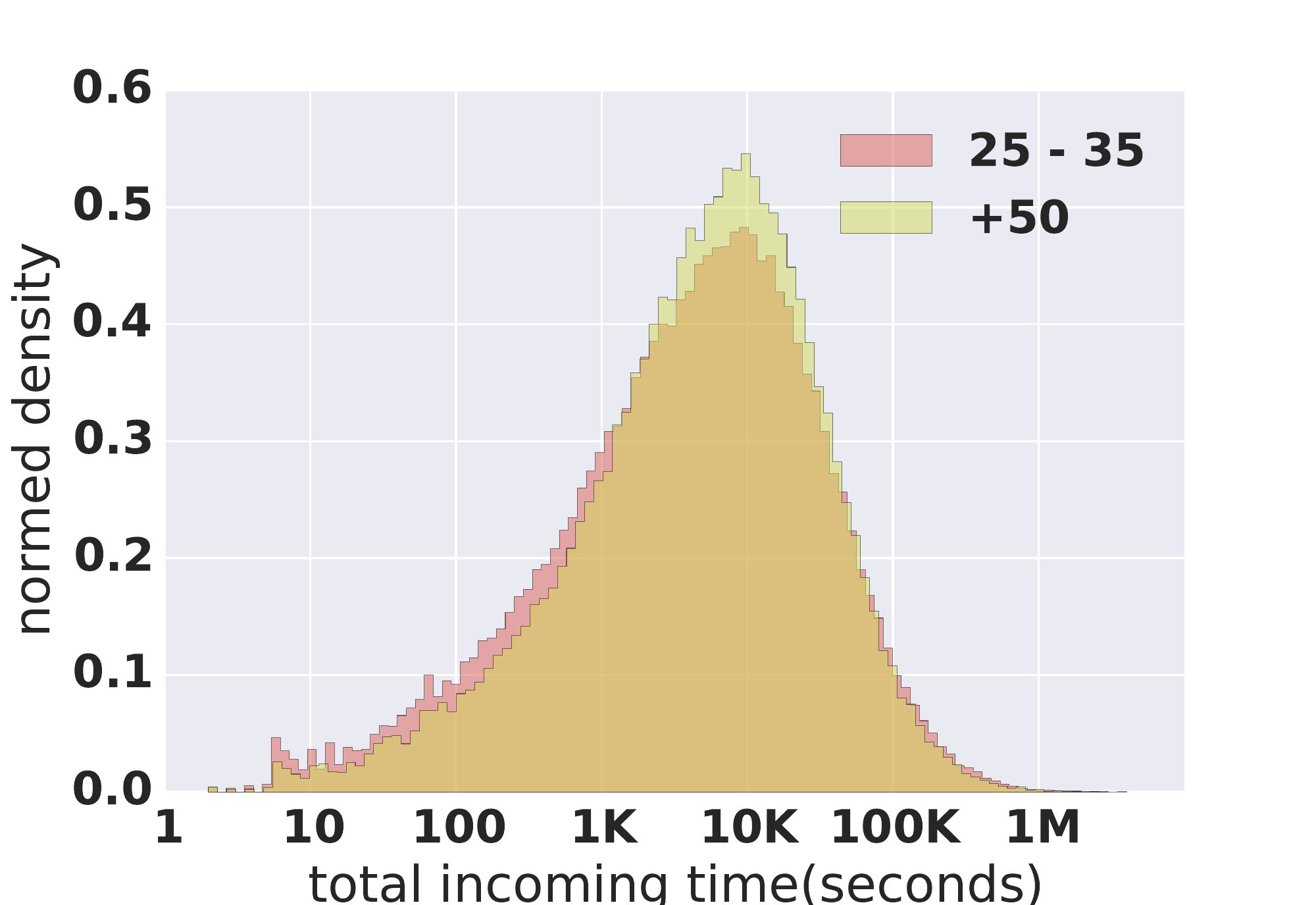}
		(e) 
	\end{minipage}
	\begin{minipage}{.48\linewidth}
		\centering
		\includegraphics[width=\textwidth]{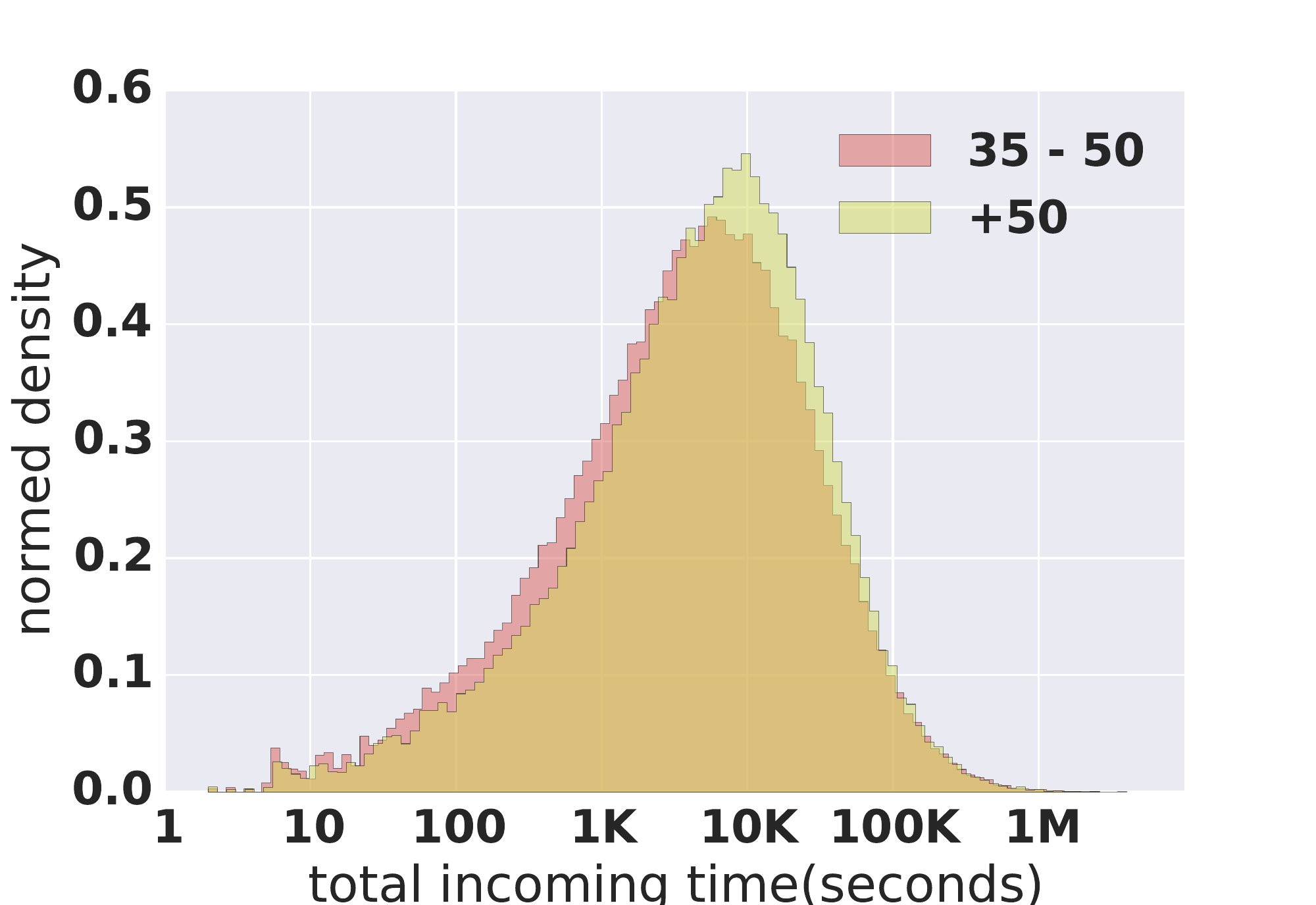}
		(f) 
	\end{minipage}
\end{footnotesize}

	\caption{Pairwise comparison of $\log_{10}(\textit{in-time-total} + 1)$  between different age groups.
	 The plots show the normalized density for $\log_{10}(\textit{in-time-total} + 1) > 0$. We observe that the distribution of the older users are most shifted to the right while the younger user's distribution is most shifted to the left, suggesting younger users tend to make shorted calls}
	\label{fig:intimetotal_comparison}
\end{figure}

We illustrate the difference between age groups for our running example: \emph{in-time-total} (total duration of incoming calls per user). 
Table~\ref{tab:tukeyHSD} shows the result of Tukey HSD (where FWER=0.05) for  
$\log_{10}$(\textit{in-time-total} + 1), obtained after the preprocessing step.
The four age groups are labeled 0, 1, 2, 3. 
Pairwise comparisons are done for all combinations (of group1 and group2). The null hypothesis is rejected for all pairs;
in other words, all the groups are found statistically different respect to this variable.

\begin{figure*}[t]
\centering
\begin{minipage}{.495\textwidth}
	\centering
	{\includegraphics[trim=1.2cm 0.2cm 2.4cm 1.2cm, clip=true, width=1.0\linewidth]
	{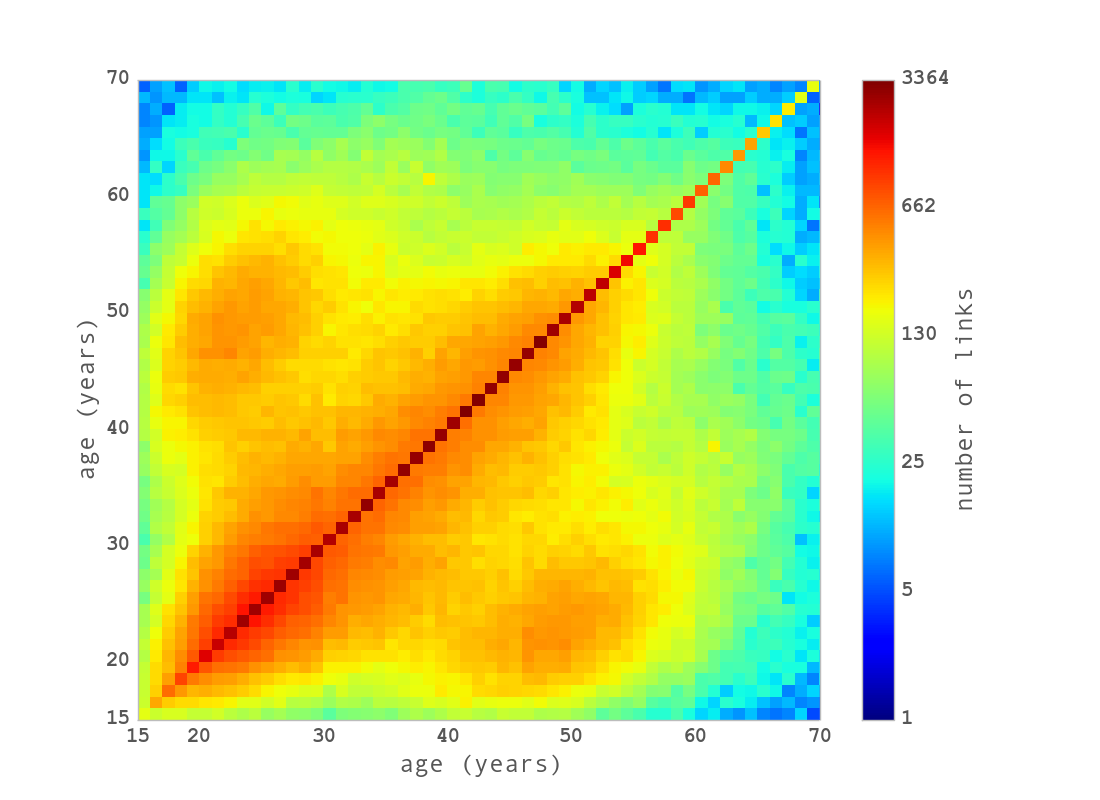}}
	(a) Communications matrix $C$.
\end{minipage}
\begin{minipage}{.495\textwidth}
	\centering
	{\includegraphics[trim=1.2cm 0.2cm 2.4cm 1.2cm, clip=true, width=1.0\linewidth]
	{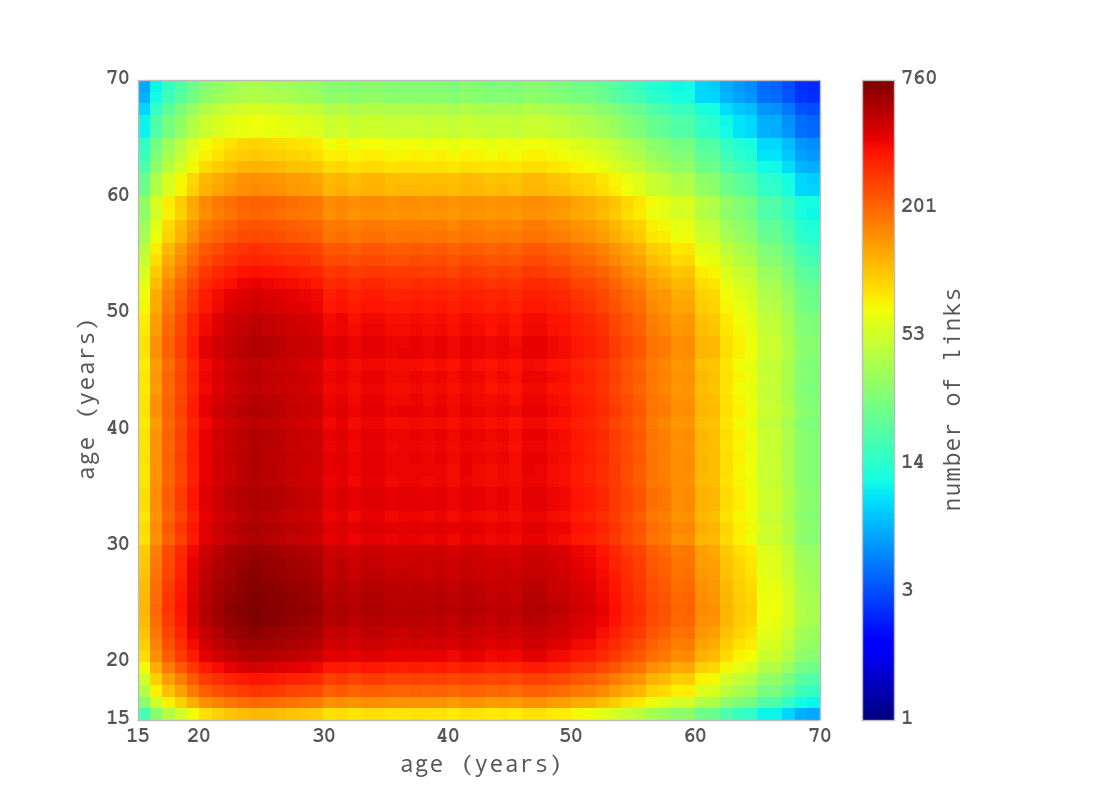}}
	(b) Random links matrix $R$.
\end{minipage}
\begin{minipage}{.495\textwidth}
	\centering
	{\includegraphics[trim=1.2cm 0.2cm 2.4cm 1.2cm, clip=true, width=1.0\linewidth]
	{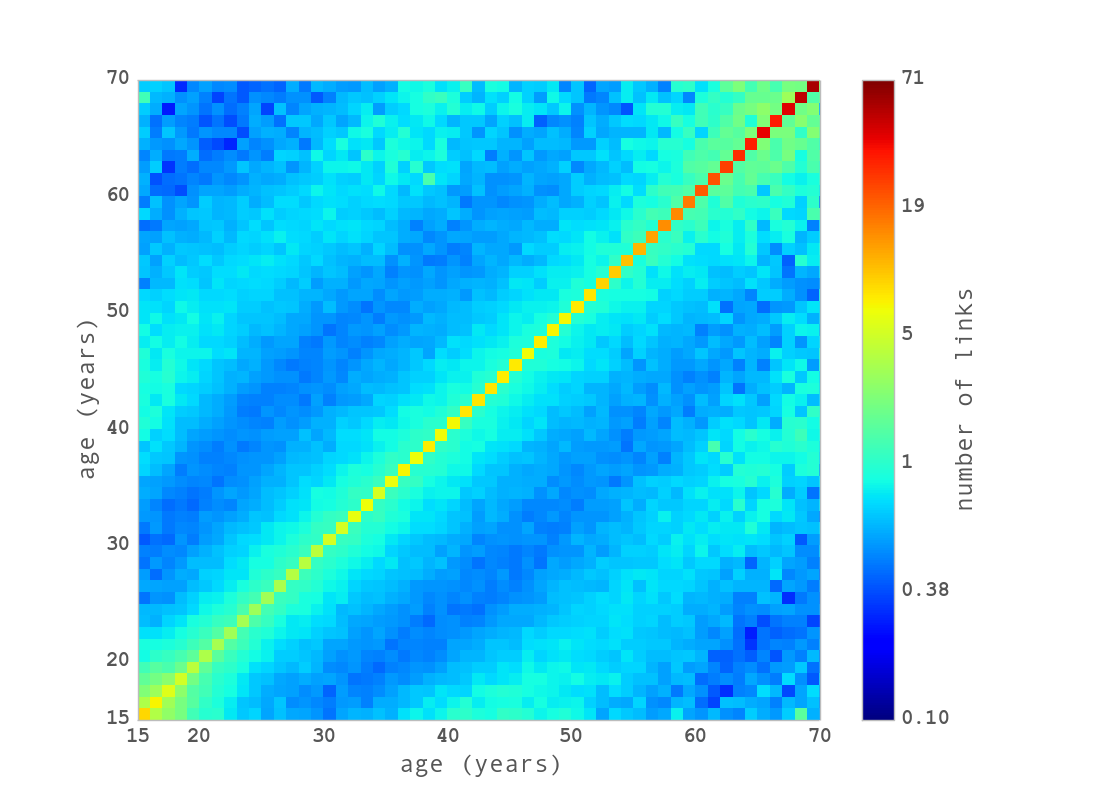}}
	(c) Difference between $C$ and $R$.
\end{minipage}
\begin{minipage}{.495\textwidth}
	\centering
	{\includegraphics[trim=1.2cm 0.2cm 2.4cm 1.2cm, clip=true, width=1.0\linewidth]
	{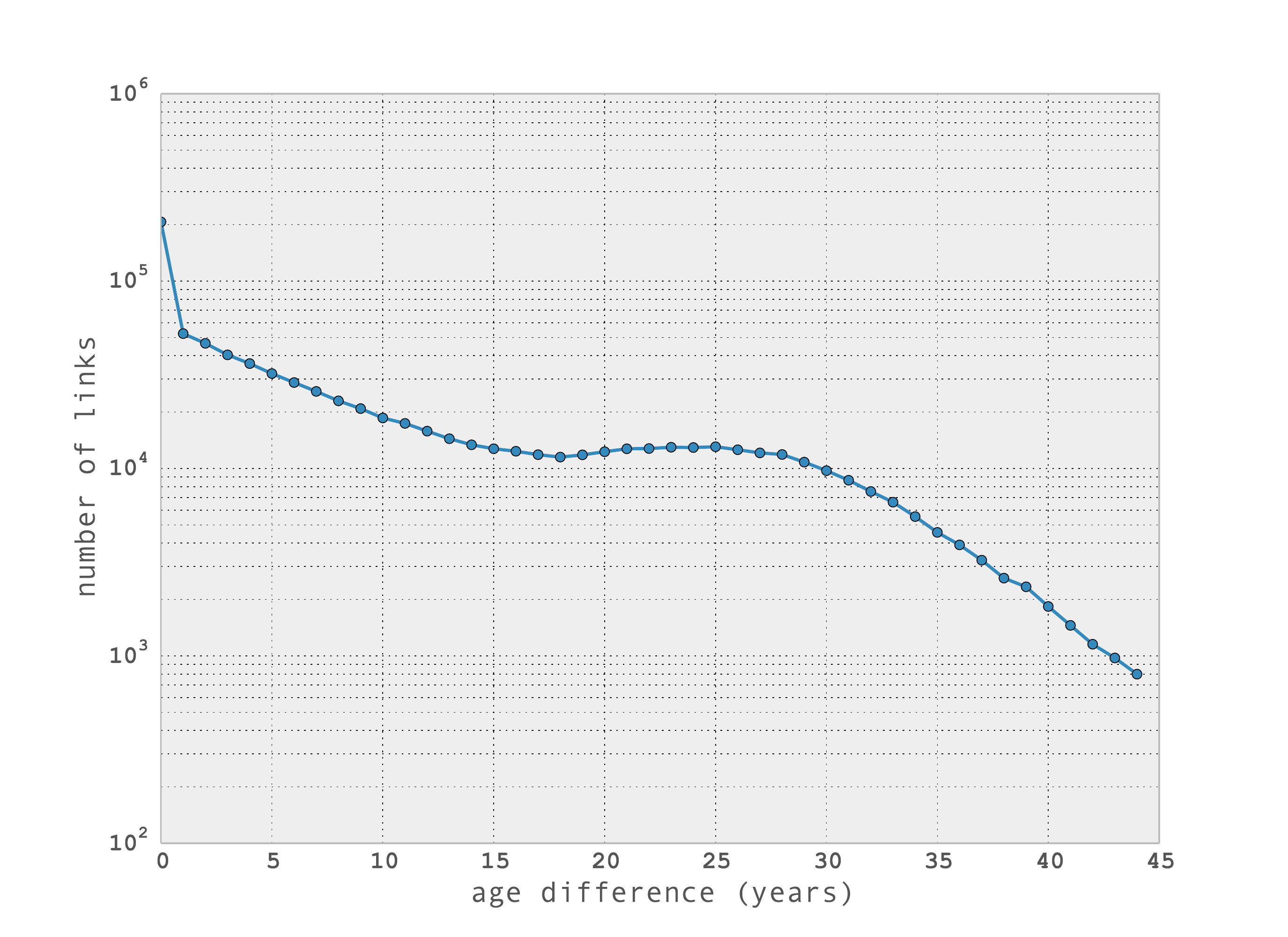}}
	(d) Number of links as function of age difference.
\end{minipage}
	\caption{Age correlation plots showing the communications matrix $C$, the random links matrix $R$, 
	the difference between $C$ and $R$, and the number of links according to the age difference. We observe strong age homophily across the whole age spectrum}
\label{fig:communications}
\end{figure*}

In Fig.~\ref{fig:intimetotal_comparison} we plot the distribution 
of $\log_{10}$(\textit{in-time-total} + 1)
for different pairs of age groups.
The following results can be observed from the plots:
\begin{itemize}
	\item The distribution for the group of people aged \textit{over 50} is shifted to the right in comparison with all the other age groups. This implies that people from this age group do talk more when called than people from any other age group. 
Figures~\ref{fig:intimetotal_comparison}(c), \ref{fig:intimetotal_comparison}(e), \ref{fig:intimetotal_comparison}(f).
	\item The distribution for the group of people aged \textit{below 25} is shifted to the left. This distribution shows less kurtosis and a higher variance, meaning 
that this population is more spread in different levels of $\log_{10}$(\textit{in-time-total} + 1). 
Figures~\ref{fig:intimetotal_comparison}(a), \ref{fig:intimetotal_comparison}(b), \ref{fig:intimetotal_comparison}(c).
\end{itemize}

\subsection{Age Homophily} \label{sec:links-age-groups}

Homophily is defined as ``the principle that contacts between similar people occurs
at a higher rate than between dissimilar people'' \citep{mcpherson2001birds}.
This basic observation about human relations can be traced back to Aristotle, who
wrote that people ``love those who are like themselves''. 
In particular, age homophily in social networks has been observed in the sociological literature,
for instance in the context of friendship in urban setting~\citep{fischer1977networks}
and other social structures~\citep{feld1982social}.
These historical studies were limited to hundreds of individuals, due to the difficulty
and cost of gathering data.
A recent study on the structure of the Facebook social graph~\citep{ugander2011anatomy}
shows that a strong age homophily can be observed in a social graph with hundreds of millions
of users.
Our dataset allowed us to verify that the age homophily principle is valid
in the mobile phone communication network on a nation wide scale.

Graph based methods like the one we present in this work rely strongly on the ability of the 
graph topology to capture correlations between node attributes we are aiming 
to predict. A most fundamental structure is that of correlations between nodes and their 
neighbourhoods. Figure~\ref{fig:communications}(a) shows the correlation matrix $\mtrx{C}$ 
where $\mtrx{C_{i,j}}$ is 
the number of links between users of age $i$ and age $j$ for the nodes in the the ground
truth $\calN_{GT}$. Though we can observe some smaller off diagonal peaks, we can see that it is 
most strongly peaked along the diagonal,
showing that users are much more likely to communicate with users of their same age. 
Performing a linear regression between the ages of linked users yields
a regression coefficient $r = 0.99$ and gradient $1.06$, confirming this observation.

To account for the population density bias shown in Fig.~\ref{fig:pyramid}, 
we compute a surrogate correlation matrix $\mtrx{R}$ 
as the expected number of edges
between ages $i$ and $j$ under the null hypothesis of independence,
plotted in Fig.~\ref{fig:communications}(b):
\begin{equation}
\mtrx{R_{i,j}} = 
\frac{ | N_{GT}(i) | }{ | \calN_{GT} | } \times
\frac{ | N_{GT}(j) | }{ | \calN_{GT} | } 
\times | \calE_{GT} | 
\end{equation} 
where $N_{GT}(i) = \{ x \in \calN_{GT} : age(x) = i \}$ 
and $\calE_{GT} = \{ (x,y) \in \calE : x \in \calN_{GT} \wedge y \in \calN_{GT} \}$.
This matrix shows the distribution of communications according to the age of nodes in a graph with the same nodes as the original, but with random edges
(while maintaining the same number of edges as the original). 
Both $\mtrx{C}$ and $\mtrx{R}$ are represented with a logarithmic color scale.
If we subtract the logarithm of $\mtrx{R}$ to the logarithm of the original matrix $\mtrx{C}$, 
we can isolate the ``social effect'' (i.e. homophily) from the pure random connections, 
as can be seen in Fig.~\ref{fig:communications}(c).

Figure~\ref{fig:communications}(d) shows the total number of links as a function of the age difference
$\delta$, showing a clear peak when the difference is $\delta = 0$.
The number of links decreases with the age difference, except around the 
value $\delta = 25$, where an interesting local maximum can be observed;
possibly relating to different generations (e.g. parents and children).
This phenomenon can also be seen in Fig.~\ref{fig:communications}(c), in 
the off diagonal bands at distance 25 years from the diagonal.

\section{Prediction using Node Attributes} \label{sec:identification}

This section describes the models that we used to estimate the age and gender of
users found in the dataset $\calN_O \setminus \calN_{GT}$ using node attributes. 
We show the results obtained using standard
Machine Learning models based on node attributes, 
applied to the prediction of gender (in Sect.~\ref{sec:gender-identification})
and age (in Sect.~\ref{sec:age-identification-node}).

We note that the feature variables are known for the whole set $\calN_O$,
while the target variables age and gender are known only for users 
in the set $\calN_{GT}$. We therefore use nodes belonging to $\calN_{GT}\cap\calN_O$
for our training and validation set, to predict both age and gender
for the remaining users in $\calN_O$.

\subsection{Population Pyramid Scaling (PPS)} \label{sec:pop-pyramid-scaling}
In the following subsections, we will present algorithms for gender and age prediction
which generate for each node, a probability vector over the possible categories
(gender or age groups).
At the end of their execution, when we ``observe'' the system, we are required to collapse
the probability vectors to a specific gender or age group. 
Choosing how to perform this collapse is not an obvious matter.
Especially in the case of age where the number of groups is greater that 2 ($C>2$).
In effect, we want to collapse the probability state of the system as a whole and not for each node independently. 
In particular, we want to impose external constraints on our solution, 
namely that the gender or age group distribution for the whole network be that of the ground truth.
To achieve this, we developed a method that we call \emph{Population Pyramid Scaling} (PPS).

More over, given $0 \leq q \leq 1$ we want our algorithm to predict over a subset of the population that represent the given fraction $q$ of the total population.
In this sense, we want the algorithm to choose the subset in a smart way, this is to predict over the subset where we are more confident about our prediction.
PPS takes, as a hyper-parameter, the proportion $q$ of
nodes to be predicted. For example, we use $q = 1/2$ to generate predictions
for the $50\%$ of nodes which got better classification results from the unconstrained method. 

\begin{algorithm}[t]
\ForEach{\emph{node $i$ and group $k$}}{
  compute the probability $p_{i,k}$ that node $i$ belongs to group $k$ 
  using the unconstrained algorithm\;
}
Create an ordered list $T$ of tuples $(i, k, p_{i,k})$\; 

Sort list $T$ in descending order by the column $p_{i,k}$.
The list $T$ will be iterated starting with the element with the highest probability\;

\ForEach{\emph{element $(i, k, p_{i,k}) \in T$}}{
		\If{\emph{node $i$ has not been assigned to a group}}{
		  \If{\emph{less than $N_k$ nodes assigned to group $k$}}{
		      assign node $i$ to group $k$\;
		  }
		}
}
\caption{Population Pyramid Scaling}
\label{algo:pps}
\end{algorithm}

The PPS procedure is described in Algorithm~\ref{algo:pps}.
We denote $N$ the size of the population to predict. 
For each category $k$ we compute the number of nodes $N_k$ 
that should be allocated to category $k$ in order to 
satisfy the distribution constraint (the gender or age distribution of $\calN_{GT}$),
and such that $\sum_{k = 1}^{C} N_k = N$ 
(where $C$ is the number of categories or groups).

For example, if $C=2$ the algorithm assigns to each category every user which probability is over a certain threshold depending on the category.
Let us remark that the algorithm fulfills the requirements about the population distribution, it is deterministic and computationally cheap.

\subsection{Gender Prediction} \label{sec:gender-identification}

For gender prediction, several algorithms were evaluated, with a preference for algorithms
more restrictive respect to the functions that they adjust.
Some of the algorithms used are:
Naive Bayes, Logistic Regression, Linear SVM, 
Linear Discriminant Analysis and Quadratic Discriminant Analysis.
As previously described, as part of data preprocessing, log transformation of the variables are added 
and the values are standardized to the $[0,1]$ interval.

\begin{table}[t]
\caption{Best classifiers configuration for Gender Prediction}
\label{tab:classifier}
\renewcommand{\arraystretch}{1.2}
\begin{tabular}{ p{0.37 \linewidth} p{0.52 \linewidth} }
\toprule
Algorithm	& Best configuration \\
\midrule
LinearSVC	& dual = True; penalty = L2;	 \\
			& loss = L1; C = 1; k = 100;		 \\
			& $|$ training set $|$ = 200,000 \\
\midrule
LogisticRegression 	& penalty = L1; C = 10; k = 100;				\\
			& $|$ training set $|$ = 200,000 \\
\bottomrule
\end{tabular}
\end{table}

The best results were obtained with Linear SVM and Logistic Regression.
Table~\ref{tab:classifier} summarizes the classifiers configuration.
To find these parameters, we used grid search over a predefined set of parameters. 
For instance, the parameter $C$ for Logistic Regression takes its values
in the set $\{ 0.1, 0.3, 1, 3, 10 \}$.
Different number of attributes were evaluated before
training the model ($k \in \{10, 30, 100\}$).
The labeled nodes were split in a training set ($70\%$) 
and a validation set ($30\%$).

\begin{table}[t]
\caption{Precision obtained for Gender Prediction. We can seed that the precession strongly improves as we select smaller subsets of users using PPS.}
\label{tab:results-gender-prediction}
\setlength{\tabcolsep}{10pt}

\begin{tabular}{  l  l l l l }
\toprule
Parameter $q$ & 1 & 1/2 & 1/4 & 1/8 \\
\midrule
Accuracy &  66.3\% & 72.9\% & 77.1\% & 81.4\% \\
\bottomrule
\end{tabular}
\end{table}

After performing PPS (to ensure the correct proportion of men and women), we obtained
the results shown in Table~\ref{tab:results-gender-prediction}.
As expected, the accuracy of our predictions improve when we decrease the parameter $q$,
which provides a trade-off between precision and coverage.
We reach a precision of 81.4\% when tagging 12.5\% of the users.

In the following paragraphs, we briefly recall some details of the classifiers that
gave the best results, in order to clarify the meaning of the configuration
parameters in Table~\ref{tab:classifier}. Those parameters are the ones required 
by the {Scikit-learn} library~\citep{scikit-learn:_2011}. 
In addition, 
{Pandas} \citep{mckinney-proc-scipy-2010} 
and {Statsmodels} \citep{statsmodels2010} 
have been used for the exploratory analysis.

\subsubsection{Linear SVC}\label{sec:linear_svc}

Classification using Support Vector Machines (SVC) requires optimizing the following 
function \citep{hsieh_dual_2008}:
\begin{equation}\label{eq:svc}
	\min_\omega \frac{1}{2}\omega^T\omega + C\sum_{i=1}^l \xi (\omega, x_i, y_i)
\end{equation}
where $y_i$ is the gender value and $x_i$ is the vector of normalized observed variables;
$\omega$ describes the hypothesis function,
and $C$ is the regularization parameter. 
In our case, we choose to use
$ \xi (\omega, x_i, y_i) = \max(1-y_i, \omega^Tx_i, 0) $.
This setup is called L1-SVM as $\xi(\cdot)$ defines the loss function.  
Implementation details can be found in 
\citep{scikit-learn:_2011,hsieh_dual_2008,fan_liblinear:_2008}.

\subsubsection{Logistic Regression}

A standard way to estimate discrete choice models is using index function models. 
We can specify the model as 
$ y^* =\omega x + \epsilon $,
where $y^*$ is an unobserved variable. 
We use the following criteria to make a choice:
\begin{align*}
\begin{cases}
	y = 1 = \text{Female}\quad & \text{if } y^* > 0\\
	y = 0 = \text{Male}\quad & \text{if } y^* < 0
\end{cases}
\end{align*}
Additionally, we use L1 regularization (for feature selection and to reduce overfitting). 
The complete  formulation is to optimize:
\[
	\min_\omega \|\omega\|_1 + C\sum_{i=1}^l \xi (\omega, x_i, y_i)
\]
where
$ \xi (\omega, x_i, y_i) = \log(1+\exp(-y_i\omega^Tx_i)) $.
For more information refer to
\citep{scikit-learn:_2011,fan_liblinear:_2008,greene_econometric_2011}.

\subsection{Age Prediction} \label{sec:age-identification-node}

We now tackle the problem of inferring the age of the users using the properties of the nodes (users). 
As a first approach, we use the Machine Learning armory to perform the detection.
We partition the target variable into $C$ age categories ($C = 4$): below 25 years old, from 25 to 34 years old, from 35 to 49 years old, and above 50 years old (as in Sect.~\ref{sec:observations-age}).
Given this set of categories, we found that Multinomial Logistic (MNLogistic) gave us
the best results in terms of precision. 
This method is a generalization of Logistic Regression for the case of multiples categories 
\citep[refer to][]{statsmodels2010,greene_econometric_2011}.

A problem we encountered when using MNLogistic is that the categories with higher frequencies
(in the ground truth) are over-represented in the predicted set; in this case the classification in categories 25 to 34 years and 35 to 49 years of more elements than expected. 
In effect, in our training set, the age groups have the distribution shown in Table~\ref{tab:training-set-distribution}.
But when using the MNLogistic algorithm, 
we obtained predictions with the distribution shown in Table~\ref{tab:mnlogistic-distribution}.

\begin{table}[b]
\caption{Age group distribution for the training set population.}
\label{tab:training-set-distribution}
\setlength{\tabcolsep}{10pt}

\begin{tabular}{  l  l l l l }
\toprule
Age group & $<25$ & 25-35 & 35-50 & $>50$ \\
\midrule
Population &  12.1\% & 35.45\% &  37.45\% & 15.0\% \\
\bottomrule
\end{tabular}
\end{table}

\begin{table}[b]
\caption{Age group distribution predicted by the multinomial logistic regression. We observe that the relative populations are significantly different from those in the training set.}
\label{tab:mnlogistic-distribution}
\setlength{\tabcolsep}{10pt}

\begin{tabular}{  l  l l l l }
\toprule
Age group &  $< 25$ & 25-35 & 35-50 & $> 50$\\
\midrule
Population & 0.66\% &  52.97\% &  45.52\% &  0.84\% \\
\bottomrule
\end{tabular}
\end{table}

To solve this issue we used the PPS (Population Pyramid Scaling) method of 
Sect.~\ref{sec:pop-pyramid-scaling}.
After performing PPS, we obtained the results presented in 
Table~\ref{tab:results-age-prediction}.

\begin{table}[t]
\caption{Precision obtained for Age Prediction based on node features using PPS. We note that PPS has constrained the relative population sized for each age group to that of the training set.}
\label{tab:results-age-prediction}
\setlength{\tabcolsep}{10.5pt}

\begin{tabular}{  l  l l l l }
\toprule
Parameter $q$ & 1 & 1/2 & 1/4 & 1/8 \\
\midrule
Accuracy &  36.9\% & 42.9\% & 48.4\% & 52.7\% \\
\bottomrule
\end{tabular}
\end{table}

\section{Prediction using Network Topology} \label{sec:age-identification-links}

As discussed in Sect.~\ref{sec:links-age-groups}, there is a strong age homophily 
among nodes in the communication network, 
yet the machine learning algorithms we have employed so far
are mostly blind to this information, that is, their predictive power relies
solely on user attributes ignoring the complex interactions given by the mobile
network they participate in.

In line with work done by \cite{Zhou04learningwith} and more generally by \cite{xu2010}, 
we propose an algorithm which can harness the information
given by the structure of the mobile network and, in this way, leverage hidden
information as the homophily patterns shown in Fig.~\ref{fig:communications}.

\subsection{Reaction-Diffusion Algorithm} \label{sec:reaction-diffusion-algorithm}

For each node $x$ in $\calG$ we define an initial state probability vector $\vect{g}_{x,0} \in \mathbb{R}^{C}$
representing the initial probability of the nodes age belonging to one of the $C = 4$ age categories 
defined in Sect.~\ref{sec:dataset}. More precisely, each component of $\vect{g}_{x,0}$ is given by
\begin{equation}
\label{eq:inititalconditions}
	(\vect{g}_{x,0})_i = \begin{cases}
		\delta_{i,a(x)} & \text{if }  x \in \calN_{S}  \\
		1/C  & \text{if } x \not\in \calN_{S} 
	\end{cases}
\end{equation}
where $\calN_{S} \subset \calN_{GT}$ is the set of seed nodes (subset of the ground truth used
for training),
$\delta_{i,a(x)}$ is the Kronecker delta function,
and $a(x)$ is the age category
assigned to each seed node $x$. For non seed nodes, equal probabilities are assigned to 
each category. These probability vectors are the set of initial conditions for the algorithm.

The evolution equations for the probability vectors $\vect{g}$ are defined as follows:
\begin{equation}
\label{eq:difusiongeneral_con_pesos}
	\begin{split}
		\vect{g}_{x,t} &= (1-\lambda) \; \vect{g}_{x,0}+ \lambda \; \frac{\sum_{x\sim y} w_{y,x} \, \vect{g}_{y,t-1} }{\sum_{x\sim y} w_{y,x}} 
	\end{split}
\end{equation} 
where $x \sim y$ is the set of $x$'s neighbours and $w_{y,x}$ is the weight of the link 
between nodes $x$ and $y$.
For our purposes, all link weights are set to one, thus the second term simplifies
to a normalized average over the values of the neighborhood nodes for $x$:
\begin{equation}
\label{eq:difusiongeneral}
		\vect{g}_{x,t} = (1-\lambda) \; \vect{g}_{x,0}+ \lambda \; \frac{\sum_{x\sim y} \vect{g}_{y,t-1} }{|\{y:x\sim y\}|} .
\end{equation}
The evolution equation is made up of two distinctive terms: the first term can be thought of 
as a memory term where in each iteration the node remembers its initial state $\vect{g}_{x,0}$; while the second term is a mean 
field term given by the average value of the neighbors 
of $x$.
The model parameter $\lambda \in [0,1]$ defines the relative importance of each of these terms. We note that a value of $\lambda=0$ implies no diffusion term while $\lambda = 1$ implies no reactive or memory term.
In our experiments, the performance of our algorithm 
was very robust to values of $\lambda$ within the $(0,1)$ range (see Sect.~\ref{sec:model-parameters}), therefore we set $\lambda=0.5$ for this study.

We note that the evolution of each term in the probability vectors are totally decoupled from one another. This last assumption in the algorithm relies on the strong homophily suggested by the correlation plots of Fig.~\ref{fig:communications}.

\paragraph{Physical Interpretation.} Next, we briefly outline our motivation for calling it 
a \emph{reaction-diffusion} algorithm.
A reaction–diffusion model explains how the concentration of one or more (typically chemical) substances changes under the influence of two processes: local reactions and diffusion 
across the medium \citep[see][]{Nicolis1977}.

Let $\mtrx{A}\in \mathbb{R}^{| \calN | \times | \calN |}$ be the adjacency matrix of $\calG$,
and $\mtrx{D}\in \mathbb{R}^{| \calN | \times | \calN |}$ the diagonal matrix with the degree of each node in $\calG$,
that is $\mtrx{A_{i,j}} = w_{i,j}$ and $\mtrx{D_{i,i}} = deg(i)$. Furthermore 
let $\mtrx{L} \in \mathbb{R}^{| \calN | \times | \calN |}$ be the matrix defined as $\mtrx{L} = \mtrx{D} - \mtrx{A}$. 
We can now rewrite equation~\eqref{eq:difusiongeneral} as
\begin{equation}
	\vect{g}_{t}^{a} = (1-\lambda) \vect{g}_{0}^{a} + \lambda 
	\mtrx{D}^{-1} [ \mtrx{D} - \mtrx{L}] \vect{g}_{t-1}^{a}
\end{equation}   
where $\vect{g}^{a}\in\mathbb{R}^{|\calN|}$  now runs over all nodes, for each age probability category $a$. We are able 
to do this since from equation~\eqref{eq:difusiongeneral}, the probabilities of each category evolve independently.
We now rewrite the above equation using the graph Laplacian for $\calG$ defined as $\mtrx{\calL}$ = $\mtrx{D}^{-1} \mtrx{L}$, also referred to as the random walk matrix:
\begin{equation}
	\vect{g}_{t}^{a} = (1-\lambda) \vect{g}_{0}^{a} - \lambda \mtrx{\calL} \vect{g}_{t-1}^{a} + \lambda \vect{g}_{t-1}^{a}
\end{equation}
finally reordering terms we get
\begin{equation}\label{eq:laplacian}
	\vect{g}_{t}^{a}-\vect{g}_{t-1}^{a} = (1-\lambda)(\vect{g}_{0}^{a}-\vect{g}_{t-1}^{a}) 
	- \lambda \mtrx{\calL} \vect{g}_{t-1}^{a}
\end{equation}
where on the left hand side we have the discrete derivative of $\vect{g}^{a}$ with respect to time, the first term on the 
right, $(1-\lambda)(\vect{g}_{0}^{a}-\vect{g}_{t-1}^{a})$, is the reactive term which drives $\vect{g}_{t}^{a}$ towards $\vect{g}_{0}^{a}$ and the second term $\lambda \mtrx{\calL} \vect{g}_{t-1}^{a}$ can be thought of as a diffusive 
term where the Laplacian operator in continuous space has been replaced by the Laplacian of the graph $\calG$ over which diffusion is taking place. From equation~\eqref{eq:laplacian}, we can see that $\lambda$ tunes the relative importance of the seed nodes information $\vect{g}_{0}$ and the topological properties of $\calG$ given by the Laplacian $\mtrx{\calL}$. We have therefore gone from a local description of the model in equation~\eqref{eq:difusiongeneral_con_pesos}, to a global description in equation~\eqref{eq:laplacian}.

Alternatively, the algorithm can be interpreted as solving a system of linear equations.
For each age category $a$ we have $g^a$, a solution to the equation:   
\begin{equation}\label{eq:sistofeq}
	(I-\lambda D^{-1}A)\vect{g}^{a}=(1-\lambda)\vect{g}_{0}^{a}.
\end{equation}
Our algorithm can thus be seen as an iterative method to solve this linear system of equations. 
It is easy to prove that the algorithm converges because 
$\rho(\lambda D^{-1}A)\leq ||\lambda D^{-1}A||_\infty=\lambda<1$
\citep[see][section 4.2]{saad2003iterative}.

Finally, let us remark that our algorithm outputs a probability vector for each node, which is very rich in information.
This is very useful because it allows us to apply PPS to impose conditions on the distribution of the population over the categories in our final prediction and to select subsets of the population for which we would expect our algorithm to work better (where the probabilities are higher).

\subsection{Reaction-Diffusion on the Mobile Phone Network}

To construct
the mobile phone network, we start with a directed edge list consisting
of $252,248,440$ edges where an edge can represent outgoing communications
from client $x$ to user $y$ and incoming communications from user $y$ to client
$x$ distinctively. This set of edges involves $93,797,349$ users.
We prune users not belonging to any connected component containing seed nodes (seed information can not diffuse to these nodes), as well as users with degree above 100, since these are most likely call centers, ending up with a 
resultant graph $\calG_T$ with  
$|\calE_T| = 125,433,585$ edges (where edges have been symmetrized) and 
$|\calN_T| = 71,949,815$ nodes, of which $|\calN_S| = 493,871$ are
seed nodes and $|\calN_V| = 164,768$ are validation nodes for which we know
their actual age group. These are used to test the inference power of our algorithm.    

As stated in equation~\eqref{eq:inititalconditions}, 
the initial conditions for each node's probability
vector $\vect{g}_{x,0}$ is that of total certainty for seed nodes and no information
for the remaining nodes in $\calG_T$. At each iteration, each
$\vect{g}_{x,t}$ in~\eqref{eq:difusiongeneral} updates its state
as a result of its initial state and the mean field resulting from the probability
vector of its neighbors. 
After the last iteration $t_{\textrm{end}}$ we assign the age group with highest probability in $\vect{g}_{x,t_{\textrm{end}}}$
to each node in $\calG_T$. Unless stated otherwise, we set the number of 
iterations to $30$ which in our experiments were sufficient for convergence 
(see Sect.~\ref{sec:model-parameters}).

\subsection{Performance by Age Group}

In Table~\ref{tb:PerformanceByAgeGroup} we present the results for the performance of the
reaction diffusion algorithm discriminating by age group.
The table shows the fraction
of correctly validated nodes for each age group ($hits$), we refer to as the performance
or inference power of the algorithm.
The age group
demographics predicted by the algorithm, both for the whole graph $\calN_T$ and
the validation set $\calN_V$, are similar to that of the seed set $\calN_S$ as can been seen in
Table~\ref{tb:demographicsfull} (the mid row refers to input data and the others to the output).
That is, the algorithm does a good job at preserving the age group demographics
of the seed set for all nodes in $\calG$.

\begin{table}[t]
\caption{Performance by age group.}
\label{tb:PerformanceByAgeGroup}
\setlength{\tabcolsep}{10pt}

\begin{tabular}{  l  r  r  r  r }
\toprule
Age group & $< 25$ & 25-34 & 35-50 & $ > 50$\\
\midrule
Performance & $26.6\%$ &  $44.6\%$ & $52.3\%$ & $45.0\%$\\
\bottomrule
\end{tabular}
\end{table}

\begin{table}[t]
\caption{Seeds and output distribution.}
\label{tb:demographicsfull}
\setlength{\tabcolsep}{9pt}

\begin{tabular}{  l  r  r  r  r }
\toprule
Age group & $< 25$ & 25-34 & 35-50 & $ > 50$\\
\midrule
Population in $\calN_T$ & $5.4\%$ &  $32.7\%$ & $37.6\%$ & $24.4\%$\\
Population in $\calN_S$ & $8.6\%$ &  $27.7\%$ & $35.9\%$ & $27.8\%$\\
Population in $\calN_V$ & $4.6\%$ &  $32.3\%$ & $35.8\%$ & $27.3\%$\\
\bottomrule
\end{tabular}
\end{table}

The age group between 35 and 50 years is significantly better predicted than the other
three age groups with a performance of 52.3\%. 
A possible  
explanation for the increased performance for the age group 35-50 can be found by
looking at the correlation plot in Fig.~\ref{fig:communications} where we observe
not only strong correlations and high population densities for this age group,
but it is also less fooled by the offside peaks mostly concentrated within the
age group 25-34. Again we see the importance of the correlation
structure among neighboring nodes. Finally, the overall performance for the
entire validation set was $46.6\%$ and we note that a performance based on random
guessing without prior information would result in an expected performance of $25\%$,
or an expected performance of $\sim 36\%$ if we set the most 
probable category (35-50) for all nodes.

\subsection{Performance Based on Topological Metrics}
\label{sec:performance-metrics}
One of the main goals of this work is to uncover topological properties 
for a given node that increase its chances of being correctly predicted, 
and characterize (based on these properties)
a subset of nodes in $\calG$ for which our algorithm works particularly well.
For this, we describe three topological metrics used to characterize each node in $\calG$: 1) the number of seeds in a node's neighborhood (\emph{SIN}), 2) the topological distance of a node to the seed set (\emph{DTS}) and 3) the degree $d$ of a node in the network, that is, the size of its neighborhood.
The main purpose of this section is to show that, given a set of seed
nodes, there are specific topological features characterizing
each node in the network that allows us to, a priori, select a set 
of nodes with optimal expected performance.  

\begin{figure}[t]
    \centering
    {\includegraphics[trim=1.5cm 0cm 1.5cm 1.0cm, clip=true, width=0.95\linewidth]
    {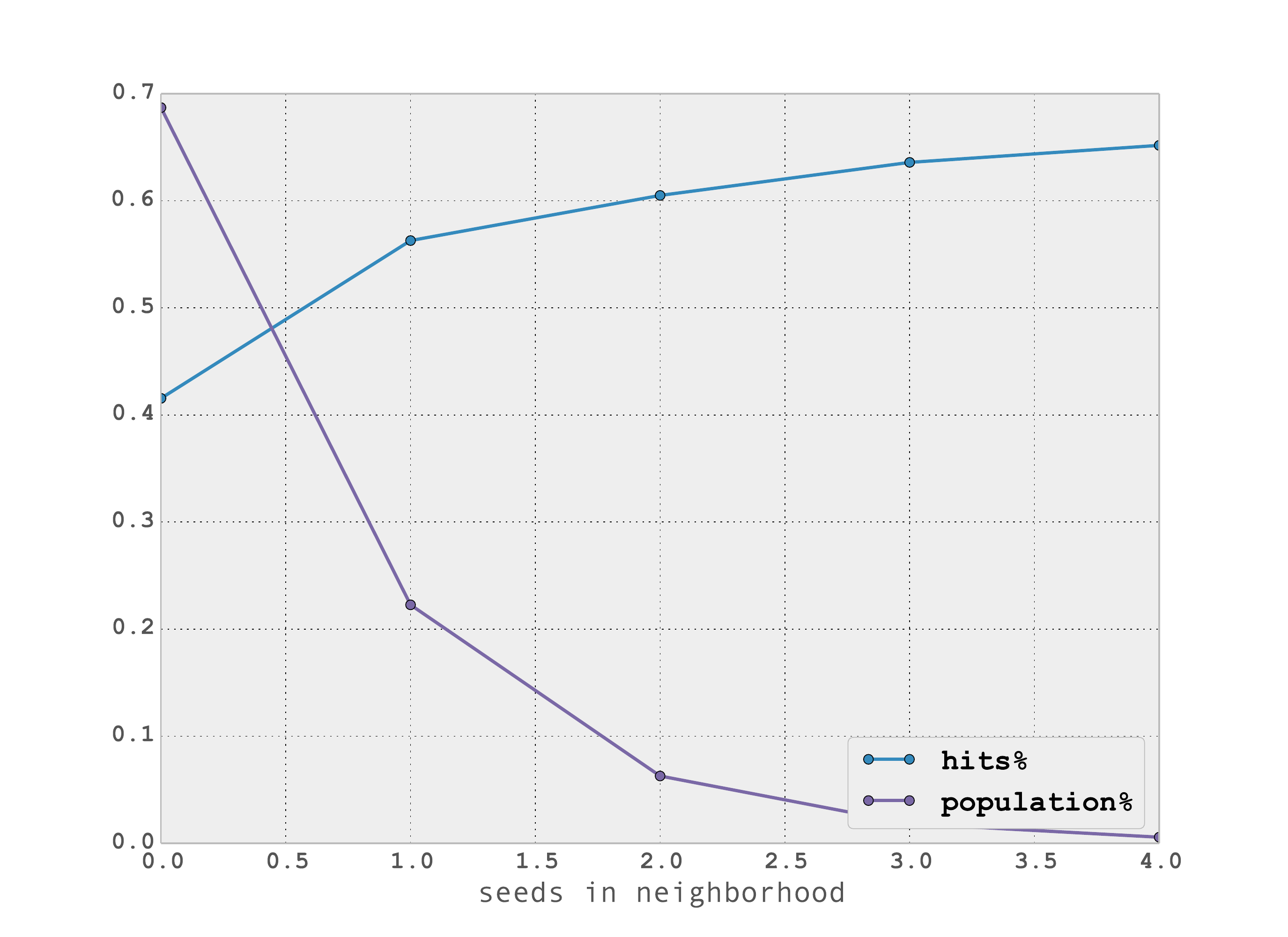}}
    \caption{Performance and population as function of \emph{SIN} (seeds in network).}
    \label{fig:seedsinego}
\end{figure}

Figure~\ref{fig:seedsinego} plots the performance (\emph{hits}) of
the algorithm over a set of nodes as a function of \emph{SIN} (number
of seeds in the nodes neighborhood). The algorithm performs
worst for nodes with no seeds in the immediate neighborhood with \emph{hits} = 41.5\%,
steadily rising as the amount of seeds increase
with a performance of \emph{hits} = 66\% for nodes with 4 seeds in their neighborhood.
We also see that the amount of nodes decreases exponentially
with the amount of seeds in their neighborhood. An interesting feature
we observed is that it is the total amount of seed nodes and not the proportion
of seed nodes that correlate most with the performance of the algorithm (this last
observation is not shown in the figure).

\begin{figure}[t]
    \centering
    {\includegraphics[trim=1.5cm 0cm 1.5cm 1.0cm, clip=true, width=0.95\linewidth]
    {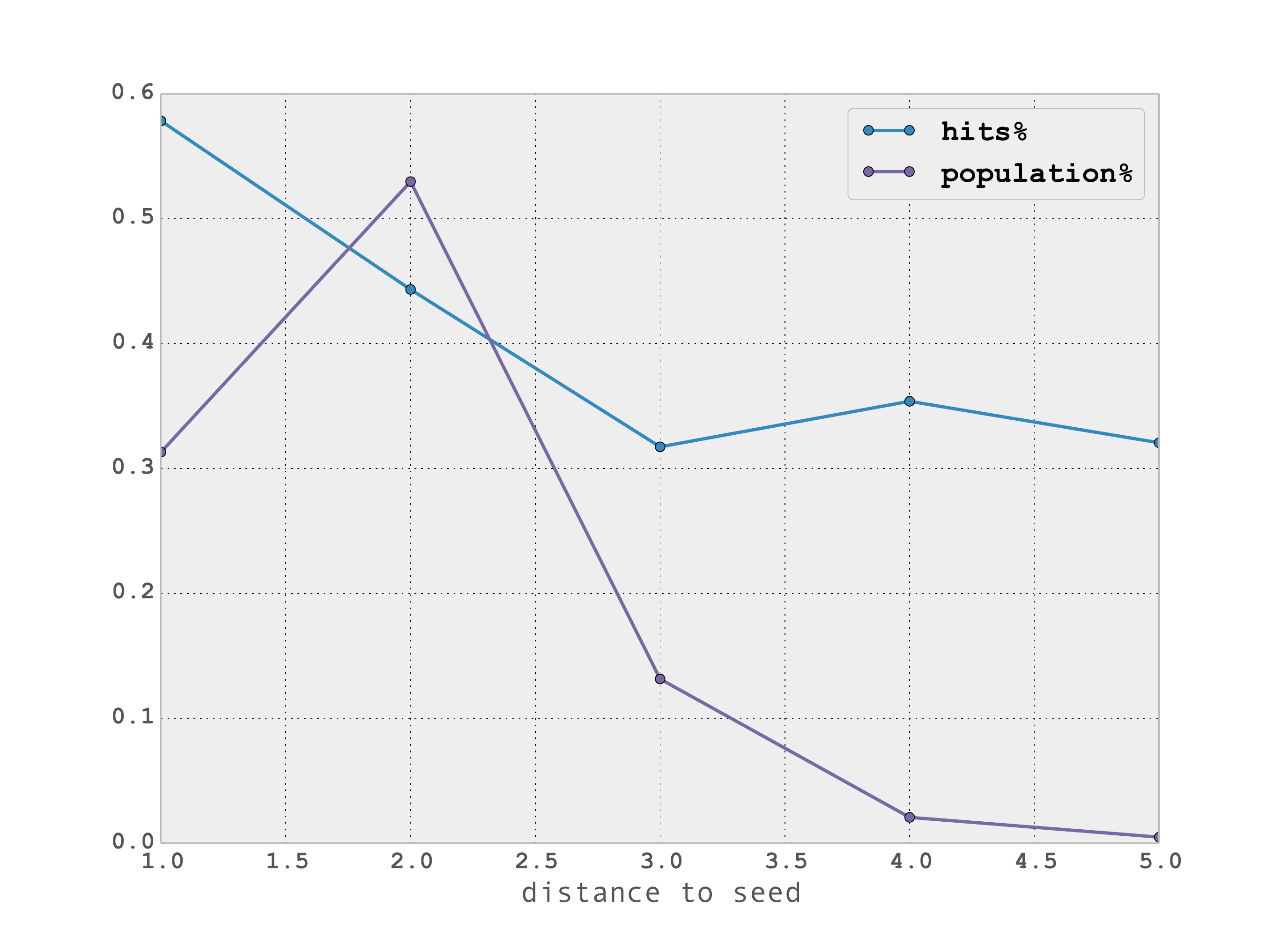}}
    \caption{Performance and population as function of \emph{DTS} (distance to seeds set).}
    \label{fig:distancetoseeds}
\end{figure}

We next examine how the algorithm performs
for nodes in $\calG_T$ that are at a given distance to the seed set (\emph{DTS}).
In Fig.~\ref{fig:distancetoseeds} we plot the population size
of nodes as a function of their \emph{DTS}. The most frequent
distance to the nearest seed is 2, and almost all nodes
are at distance less than 4. This implies
that after four iterations of the algorithm, the seeds information
have spread to most of the nodes in $\calG_T$. This figure also shows
that the performance of the algorithm decreases as the distance of a node
to $\calN_S$ increases.

Both \emph{SIN} and \emph{DTS}, although topological quantities themselves, are defined in terms of
the distribution of the seeds set, and are not intrinsic properties the graph $\calG_T$. 
We therefore look at a basic intrinsic topological property of a node
$x$ in $\calG_T$, namely its degree $d(x)$.

\begin{figure}[t]
	\centering
    {\includegraphics[trim=1.7cm 0cm 2.0cm 1.0cm, clip=true, width=0.95\linewidth]
	{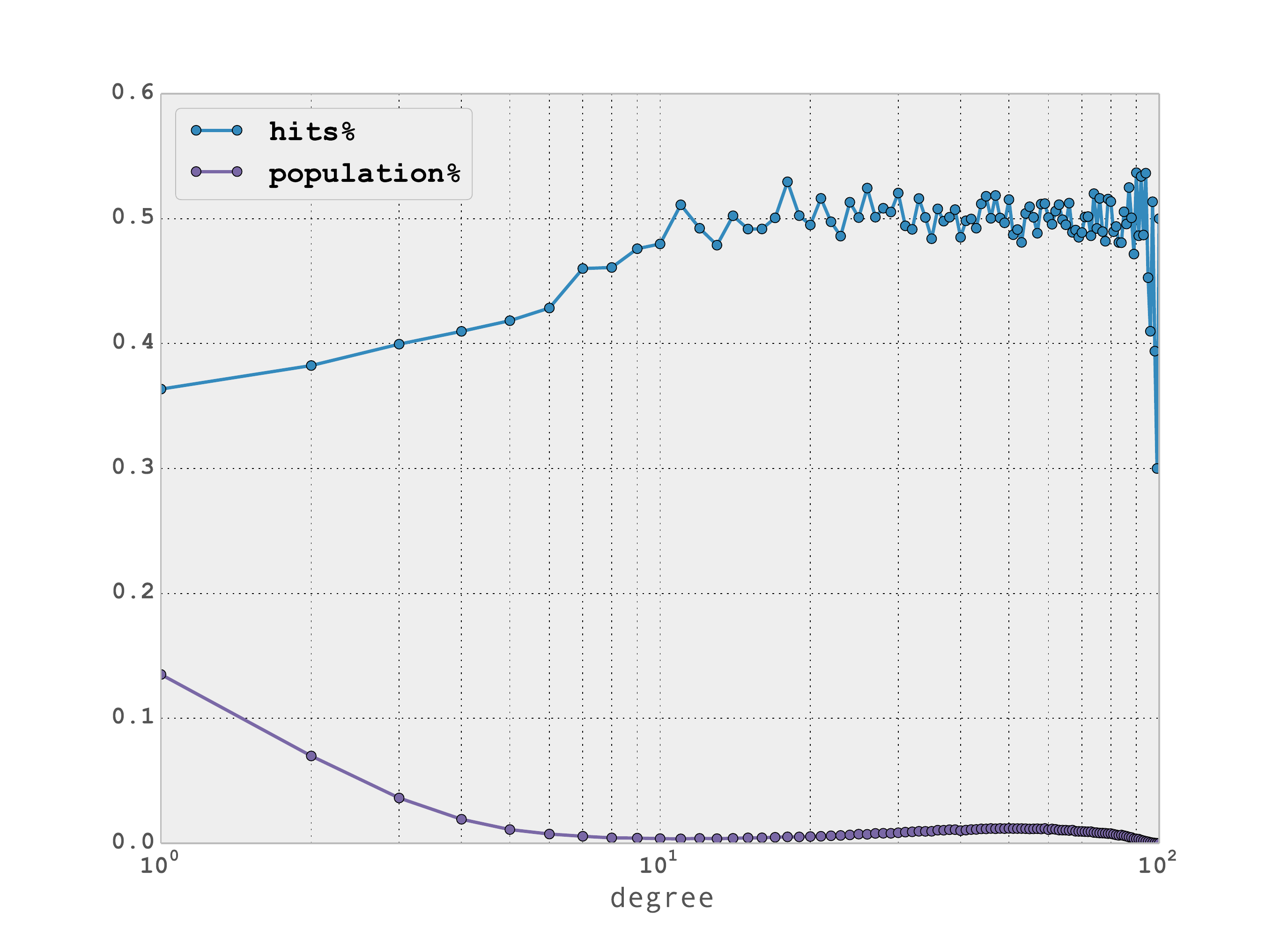}}
	\caption{Performance and population as function of the nodes degree.}
	\label{fig:nodesdegree}
\end{figure} 

In Fig.~\ref{fig:nodesdegree} we see that the performance
of the algorithm is lowest for nodes with small degree and gradually increases as
the degree increases reaching a plateau for nodes with $d(x) > 10$. 

Understanding what makes degree
values close to 10 an inflection region in the performance of the reaction-diffusion
algorithm is something we will address in a future work, but it is here worth remarking
that $\calG_T$ is an essentially incomplete graph, where we are missing all the communication
links among users that are not clients of the mobile operator. An important topological
consequence of this is that whereas the average degree over the whole graph $\calG_T$
is $3.48$, the average degree for client nodes is $37.53$.

\begin{table}[t]
\caption{Performance as function of degree and distance to seeds.}
\label{tb:optimalmetrics}
\setlength{\tabcolsep}{12pt}

\begin{tabular}{ l c c c } 
\toprule
& \multicolumn{3}{c}{Distance to Seeds} \\ \cmidrule(l){2-4}
Degree & 1 & 2 & 3 \\  
\midrule
{[}1, 2] & 0.415057 & \textbf{0.465881} & 0.313735 \\ 
(2, 29] & 0.581540 & 0.445736 & \textbf{0.334968} \\ 
(29, 48] & \textbf{0.618067} & 0.435747 & 0.299578 \\ 
(48, 66] & 0.588592 & 0.438933 & 0.288136 \\ 
(66, 100] & 0.566038 & 0.441583 & 0.206897 \\ 
\bottomrule
\end{tabular}
\end{table}

Having shown the strong correlations between the three topological
metrics and the algorithm's performance -- which allows us to select
optimal values for each metric independently -- we next examine whether
we can find an optimal combination of these metrics.
Table~\ref{tb:optimalmetrics} shows the performance of our algorithm for different values 
of their \emph{DTS} and their degree. 
Even though we have shown that
performance increases with smaller distance to seeds and higher degree, here we find
that this rule can be violated for specific ranges of the nodes degree -- as is shown
for nodes with degree between 1 and 2
where performance is optimal for nodes at distance 2 from the seed set. 
Most interestingly, we find that selecting a subset based
only on the nodes degree and distance to seeds, the optimal set is at distance 1
from the seed set and with degree between 29 and 48 for which
the performance  of the algorithm increases to \emph{hits} = 61.8\% with 20,050 nodes in $\calN_V$
satisfying these set of conditions. This last result indicates that it is not sufficient
to look at each metric independently to select the subset of nodes for optimal performance
of the algorithm. We also looked at the performance of the algorithms as a function of \emph{SIN} and 
degree, and also found subsets of nodes for whom performance increased to similar values.

\subsection{Applying Population Pyramid Scaling} \label{sec:pyramidscaling}

A salient feature of our algorithm is that the demographic information being spread 
is not the age group itself, but the probability vector of the age groups.
In each iteration, the algorithm does not collapse the information in each node
to a preferred value; instead it allows the system to evolve as a probability state over the network.
As we mentioned in Sect.~\ref{sec:pop-pyramid-scaling}, 
to generate the final output, we collapse
the probability vectors to specific age groups; and
choosing how to perform this collapse is not an obvious matter.
In the previous sections we selected for each node the category with highest probability.
This is a natural approach, 
and as we have shown, the algorithm's performance is satisfactory. 
Furthermore, the problem of over-representation of most common categories reflected in Table~\ref{tab:mnlogistic-distribution} was not observed in this case.
In order to have a natural and clean way to compare results between the different approaches,
we apply PPS with the \emph{reaction-diffusion} algorithm as well.
Table~\ref{tab:results} (column RDif) shows the results that we obtained.

\subsection{Combined Algorithm Performance} \label{sec:results-summary}

We describe here an algorithm to predict the age of users that leverages 
the \emph{PPS} algorithm, the node classification of section~\ref{sec:age-identification-node}
and the pure graph-based \emph{Reaction-Diffusion} algorithm. 
We define as initial state:
\begin{equation}
	f_{x} = \begin{cases}
		\delta_{i,a(x)} & \text{if }  x \in \calN_{S}  \\
		\text{ML}(x)   & \text{if } x \not\in \calN_{S}
	\end{cases}
\end{equation}
where $\text{ML}(x)$ is the result given by the best \textit{Machine Learning} algorithm of section~\ref{sec:age-identification-node} 
(i.e. Multinomial Logistic),
and $\calN_{S}$ is the set of seed nodes. 
Then, as before, the iterative process follows equation~\eqref{eq:difusiongeneral}.
In this case, the hyper-parameter $\lambda$ provides a trade-off between the information from the network topology and the initial information obtained with Machine Learning methods over node attributes
(here again we take $\lambda = 0.5$).

Table~\ref{tab:results} summarizes the results obtained with the different methods:
Machine Learning (ML) alone, Reaction-Diffusion (RDif) alone, and the combined method (ML + RDif).
We report for each case the accuracy obtained, that is the percentage of correct 
predictions on the validation set.

\begin{table} [t]
\caption{Precision obtained for Age Prediction}
\label{tab:results}
\setlength{\tabcolsep}{13pt}

\begin{tabular}{ c c c c }
\toprule
\textbf{Population} & \textbf{ML} & \textbf{RDif} & \textbf{ML + RDif}\\
\midrule
$q$=1~~ & 36.9\% & 43.4\% & 38.1\%	\\
$q$=1/2 & 42.9\% & 47.2\% & 46.3\%	\\
$q$=1/4 & 48.4\% & 56.1\% & 52.3\%	\\
$q$=1/8 & 52.7\% & 62.3\% & 57.2\%\\
\bottomrule
\end{tabular}
\end{table}

The table shows that taking a smaller $q$ improves the accuracy of the results.
Our experiments also show that the RDif (Reaction-Diffusion) algorithm outperforms 
the ML predictions based on node attributes.
It is also interesting to remark that the RDif algorithm outperforms
the combined method. 
The best precision obtained is 62.3\% of correctly predicted nodes, 
when tagging 12.5\% of the population.
Note that random guessing the age group (between 4 categories) would yield a precision of 25\%,
and a random prediction leveraging the population distribution, according to Table~\ref{tab:training-set-distribution}, would give a precision of 30.3\%.

\subsection{Sensitivity to Model Parameters} \label{sec:model-parameters}
So far we have analyzed the performance of the reaction-diffusion algorithm
without addressing the issue of model parameters selection, namely the 
parameter $\lambda$ presented in Sect.~\ref{sec:reaction-diffusion-algorithm}
-- which tunes the relative importance of the reaction and diffusion terms --
together with the number of iterations 
of the algorithm, which determines to which extent the age information in the seed set
spreads across the network. 
In this section we examine these two parameters for the client network $\calG_C$
consisting only of client nodes.
This graph is made up of $| \calE_C | = 6,484,571$ edges 
and $| \calN_C | = 3,225,538 $ nodes of which 349,542 nodes  
are seed nodes and 143,240 nodes are validation nodes.

\medskip
\noindent
\emph{Parameter $\lambda$.}
For $\lambda=0.0$ (no diffusion) we have $\vect{g}_{x,t} = \vect{g}_{x,0}$ for all $t$.
As the diffusion term is slightly turned on, $\lambda=10^{-7}$, 
the algorithm gets a performance of $48.3\%$ and remains almost constant for $0 < \lambda < 1$.
It drops suddenly to a performance of $42.5\%$ for $\lambda=1.0$ 
(total diffusion with no reactive term) and we can't ensure convergence in this case.
We conclude from these results that inside the boundary values
for $\lambda$ the performance remains almost unchanged and thus the algorithm is very robust to parameter
perturbations.
In other words, our algorithm needs some diffusion and reactive terms but the relative
weights of these are unimportant.    
Finally, optimally choosing the reactive term, we can get a performance of $48.8\%$.

\medskip
\noindent
\emph{Convergence.}
An important property of the model we address here is that of the model's time scale, 
that is, how long it takes the system to reach a stationary state. 
Stationarity analysis in complex networks such as the one we study here is a hard problem, thus for the purpose of this work we limit ourselves to the notion of performance.
We therefore say the algorithm reaches a stationary state after $t$ iterations 
if its performance remains unchanged for further iterations. 
We have seen that most nodes in the graph are not directly connected to a seed node,
and that the diffusion term in Equation~\eqref{eq:difusiongeneral} spreads from neighbor to neighbor
one iteration at a time.
We therefore expect that as iterations increase, the seed nodes'
information will spread further across the network. 
Our experiments show that after
a single iteration, the performance of the reaction-diffusion algorithm is already at $45.5\%$ gradually increasing
its performance to a stationary value of $\sim 50\%$ after 5 iterations.
As conclusion, setting the number of iterations as $t_{\textrm{end}} = 30$ gives us a good margin to ensure convergence
of the algorithm.

\section{Related Work} \label{relatedwork}
The use of mobile phones as surrogates for the study of human dynamics and social interactions has attracted significant attention from the research and industry community in recent years,
as shown by the survey of \cite{naboulsi2015mobile}. 
\citet{blumenstock2010mobile} and \citet{blumenstock2010s} performed a comparative study between the demographics of mobile phone users and that of the general population of Rwanda -- showing significant differences for the demographics of gender and age of the mobile network population compared to the population of Rwanda at large. 
In our study, we also observed significant differences between the demographics of age and gender population of mobile phone users and that given by the national census. 
\citet{blumenstock2010s} also found differences between men and women similar to the ones we report in Table~\ref{tab:mean-male-female}.
Mobile phones datasets have also been used to study other demographic attributes. \citet{GutierrezKringBlondel2013} used the air time credit purchases of a mobile phone in Ivory Coast 
to estimate the socio-economic status of different geographic areas of the country. 
Using CDRs 
they where also able to show significant cohesion in the amount of air time purchases within network communities.   

Significant research has focused on the structure of mobile phone networks, in order to better understand the social implications of these properties. \citet{OnnelaPNAS2007} looked at the relation between the strength of communication ties between mobile phone users, and showed that weak ties are most important in keeping the network from falling apart, verifying the importance of weak ties in a mobile network. 
Power laws have usually been the standard statistical model for many network variables, but in \citep{SeshadriMachirajuFaloutsos2008} a careful study of the number of calls, duration of calls and number of calling pairs per user in a mobile phone network showed significant deviations from these models and instead were better modelled by a double Pareto lognormal distribution. In contrast, the extensive work done for the Facebook friendship network in \citep{ugander2011anatomy} showed that the number of friends per user distributions can be successfully modelled as a power law. In our work we have not tried to rigorously model these distributions but we did observe that the calling pairs distribution can be roughly modelled as a power law. Furthermore, most social networks are intrinsically dynamic and time-scales of these changes are important for many sociological studies \citep{BarratVespiniani2008}. In particular, \citet{MiritelloMoro2013} show how in a mobile phone network the local topological properties of a node (such as being a hub) can significantly affect its sensitivity to dynamic changes.    

\citet{DongTangChawla2013} studied the relationship between call duration and social ties, showing that pairs with strong ties make more but shorter calls. They use this study to develop a time dependent factor model to infer time duration of calls. Recently, \citet{DongTangChawla2014} used the log records of a  mobile phone network in India to predict the age and gender of mobile phone users. In their data set they observed strong correlations between the age and the gender of the callers. This allowed them to propose a novel  approach using factor models where inference of age and gender was carried out simultaneously. It is not clear whether this strategy would work on our dataset given that the homophily between gender groups was not very strong, but it is an idea 
that we are planning to study. 

In \citet{sarraute2014} we provided an observational study of differences in mobile phone usage according to gender and age using CDRs.
We set out to infer the age categories and gender of the users, by applying several standard machine learning tools. The strong age homophily in the network motivated us to propose an algorithm based on the network topology to improve our predictive performance on the age categories of users.
We proposed a diffusion algorithm where the probability of each age category is diffused across the network. This strategy of having an information vector of possible values of the target variable diffusing across the network, instead of selecting a specific value at each iteration step, has also been proposed by \citep{Zhou04learningwith} and \citep{zhou2004semi}, and has been shown to be a specific instantiation of the larger framework of Krigging as described by \cite{xu2010}. In our particular implementation, the information vector is given by a probability vector, which is iterated with a probability matrix. This allows us to preserve the probability state for each node until the last iteration. We then provide a method to collapse the probability vectors as a whole and not for each node independently. Selecting the age categories for each node based on the outcome probabilities for all nodes in the network allows for us to impose further global constraints on the inferred results.

In the present work, we further exploit the properties of the network topology, by analyzing how the topological properties of nodes and their relation to the seed nodes can influence our capacity to infer the age category.
We found that having a seed in a user's neighborhood can significantly increase our predictive power for that user, but only a further  marginal increase in performance is found if the neighborhood has more that one seed -- i.e. one seed in a user's neighborhood is enough for a significant increase in performance. 
Another extension of this work was to  study the robustness of the \emph{reaction-diffusion} algorithm to the parameter $\lambda$: we showed that it is very robust to the whole set of values $(\epsilon,1-\epsilon)$ where $\epsilon>0$, i.e. 
the algorithm requires to have some reactive term and some diffusive term, but their relative weight is not important.  
We also deepened the observational study analyzing in more details the difference between genders and age groups. 

\section{Conclusion and Future Work} \label{sec:conclusion}

Large scale social networks such as those emerging from mobile phone communications are
rapidly increasing in size and becoming more ubiquitous around the globe. Understanding
how these community structures forming complex network topologies can inform mobile
phone operators, as well as other organizations, on unknown attributes of their users is of
vital importance if they are to better understand not only their clients' interests and behavior, but
also their social environment including the
users that are not clients of the mobile operator. A better description of this
ecosystem can give the mobile operators a business edge in areas such as churn prediction,
targeted marketing, and better client service among other benefits.

This work extends the study of
social interactions focusing on gender and
age, made by \citet{sarraute2014}.
From a sociological perspective, the ability to analyze
the communications between tens of millions of people allows us to make strong inferences
and detect subtle properties of the social network.

As described in Sect.~\ref{sec:exploratory}, the graph we constructed has
very rich link semantics, containing a detailed description of the communication patterns.
With PCA, we found that most of the variance of the characterization variables is contained in
a low dimensional subspace generated by vectors with clear semantic interpretation.
Motivated by these results, we focused on how the statistical properties of
the most informative attributes vary with both gender and age. 
In Sect.~\ref{sec:observations}, 
we make two interesting observations: 
(i) there is a gender homophily in the communication network;
(ii) an asymmetry respect to incoming and outgoing calls can be observed between men and women,
possibly reflecting a difference of roles in the studied society.
Our most important observation is the study of correlations
between age groups in the communication network, as summarized in Fig.~\ref{fig:communications}.
We observe a strong age homophily,
and a strong concentration of communications centered 
around the age interval between 25 and 45 years. 
But we also notice weaker modes in both figures,
which raise interesting sociological questions
(e.g. whether they reflect a generational gap).

The second key contribution of this work is to propose and analyze  methods
to infer the age of users in the mobile network.
As a first approach, described in Sect.~\ref{sec:gender-identification} and \ref{sec:age-identification-node}, we used a set of standard Machine Learning tools based on node attributes.
However, these techniques cannot harness the topological information
of the network to explore possible correlations between the users' age groups. 
To leverage this information, we proposed a purely graph-based algorithm inspired in a 
\emph{reaction-diffusion} process, and demonstrated that with this methodology we could 
predict the age category for a significant set of nodes in the network.

In our analysis, we focused on the bare bones
topology of the mobile network: in other words, we aimed at uncovering the potential of the topology
of the network itself to inform us on user attributes, in particular on
the users age group.
The only prior knowledge for our proposed inference algorithm is
the age group of a subset of the network nodes (the seed nodes), and the
network topology itself. 

We next searched for basic
topological features to guide us in finding a subset of nodes where our
algorithm performs best.
Optimizing over all three metrics described in Sect.~\ref{sec:performance-metrics},
we found a subset of nodes where the performance of our algorithm increased 
to $62\%$ -- whereas pure random selection would have achieved a performance of $25\%$.

In conclusion, in this work we have presented an algorithm that can harness 
the bare bones topology of mobile phone
networks to infer with significant accuracy the age group of the network's users.
We have shown the importance
of understanding nodes topological properties, in particular their relation to the seed nodes,  
in order to fine grain our expectation of correctly classifying the nodes.
Though we have carried out this analysis for a specific network using a particular algorithm, we believe this approach can be useful to study graph-based prediction algorithms in general.

\medskip
\noindent 
There are multiple directions in which this work can be extended.
We highlight the following:

\emph{Extend Depth.}
A statistics quasi-experiment can be built from this method \citep{william2002experimental}.
In this case, we want to know whether the differences in the observed behavior can be accounted 
to gender and age, or are consequences of differences in the ego-network induced by phone calls. 
This quasi-experiment can be performed using Propensity Score \citep{rosenbaum1983central},
and may provide sociological insights.

\emph{Extend Width.}
One direction that we are currently investigating is to apply the methodology presented here
to predict variables related to the users' spending behavior.
In \citep{singh2013predicting} the authors show correlations between social features
and spending characterizations, for a small population (52 individuals).
We are interested in applying our methodology to predict spending behavior characteristics
on a much larger scale (millions of users).

\emph{Mobility.}
Another research direction is to use the geolocation information contained in the Call Details Records.
Recent studies have focused on the mobility patterns related to cultural events
-- for instance sport related events \citep{ponieman2013human,xavier2013understanding} --
which might exhibit differences between genders and age groups.
Looking at mobility patterns through the lens of gender and age characterization will provide
new features to feed the Machine Learning part of our methodology,
and more generally will provide new insights on the human dynamics of different segments of the population.

\bibliographystyle{plainnat}
\balance
\bibliography{../bibliography/sna}

\end{document}